\documentclass[letterpaper,twocolumn,10pt]{article}
\usepackage{hyperref}
\hypersetup{
    colorlinks,
    linkcolor={red!50!black},
    citecolor={blue!50!black},
    urlcolor={blue!80!black}
}
\usepackage{usenix-2020-09}

\usepackage{tikz}
\usepackage{amsmath}
\usepackage{algorithm}
\usepackage{algpseudocode}
\usepackage{backnaur}
\usepackage{amsmath,amsfonts}
\usepackage{caption}
\usepackage{subcaption}
\usepackage[acronym]{glossaries}
\usepackage{xspace}
\usepackage{tabularx}
\usepackage{xcolor}
\usepackage{url}
\usepackage{tikz}
\usepackage{listings}
\usepackage{graphicx}
\usepackage{multirow}
\usepackage{comment}
\usepackage[frozencache,cachedir=minted-cache]{minted}
\usepackage{paralist}
\usepackage{booktabs}

\usepackage{jade_common}

\newcommand{\name}{CensorLab\xspace}

\newcommand{\xmark}{%
{\color{red}{
\tikz[scale=0.23] {
    \draw[line width=0.7,line cap=round] (0,0) to [bend left=6] (1,1);
    \draw[line width=0.7,line cap=round] (0.2,0.95) to [bend right=3] (0.8,0.05);
}}}}
\newcommand{\cmark}{%
{\color{green}{\tikz[scale=0.23] {
    \draw[line width=0.7,line cap=round] (0.25,0) to [bend left=10] (1,1);
    \draw[line width=0.8,line cap=round] (0,0.35) to [bend right=1] (0.23,0);
}}}}
\newcommand{\maybemark}{%
{\color{orange}{\normalsize ?
}}}
\newcommand{\maybecheck}{%
{\color{orange}{\tikz[scale=0.23] {
    \draw[line width=0.7,line cap=round] (0.25,0) to [bend left=10] (1,1);
    \draw[line width=0.8,line cap=round] (0,0.35) to [bend right=1] (0.23,0);
}}}}

\date{}

\title{\Large \bf CensorLab: A Testbed for Censorship Experiments}

\author{
{\rm Jade Sheffey}\\
University of Massachusetts Amherst
\and
{\rm Amir Houmansadr}\\
University of Massachusetts Amherst
} 
\begin{document}
\maketitle

\begin{abstract}
Censorship and censorship circumvention are closely connected, and each is constantly making decisions in reaction to the other.
When censors deploy a new Internet censorship technique, the anti-censorship community scrambles to find and develop circumvention strategies against the censor's new strategy, i.e., by targeting and exploiting specific vulnerabilities in the new censorship mechanism. 
%
We believe that over-reliance on such a \emph{reactive} approach to circumvention has given the censors the upper hand in the censorship arms race, becoming a key reason for the inefficacy of in-the-wild circumvention systems. 
Therefore, we argue for a \emph{proactive} approach to censorship research: 
the anti-censorship community should be able to proactively develop circumvention mechanisms against hypothetical or futuristic censorship strategies.

To facilitate proactive censorship research, we design and implement \name, a generic platform for emulating Internet censorship scenarios. \name aims to complement currently reactive circumvention research  by efficiently emulating past, present, and hypothetical censorship strategies in  realistic network environments. Specifically,  \name aims to (1) support all censorship mechanisms previously or currently deployed by real-world censors; (2) support  the emulation of hypothetical (not-yet-deployed) censorship strategies including advanced data-driven censorship mechanisms  (e.g., ML-based traffic classifiers); (3) provide an easy-to-use platform for researchers and practitioners enabling them to perform extensive experimentation; and (4) operate efficiently with minimal overhead.
We have implemented \name as a fully functional, flexible, and high-performance platform, and showcase how it can be used to emulate a wide range of censorship scenarios, from traditional IP blocking and keyword filtering to hypothetical ML-based censorship mechanisms. 
\end{abstract}


\section{Introduction}


Internet censorship poses a significant threat to the free flow of information across the globe, and is considered by many as a serious violation of human rights~\cite{internet-human-rights}. Unfortunately, censorship has become both increasingly prevalent and potent in recent years. The Great Firewall of China (GFW) has deployed novel, sophisticated techniques to censor encrypted tunneling protocols~\cite{Wu2023a, 10.1145/3419394.3423644}, while Iran has deployed drastic network censorship measures with little regard towards collateral damage~\cite{geneva-Bock2019a}. Furthermore, Russia~\cite{russia-censorship-xue2022b} and India~\cite{india-censorship-Katira2023a} have also implemented wide-reaching censorship apparatuses.

In response, the Internet freedom community has been developing various censorship \emph{circumvention} mechanisms~\cite{geneva-Bock2019a, domainfronting-fifield2015blocking, Sharma2021a, Lorimer2021a} aimed to bypass censorship and provide users in censored regions the means to access the otherwise restricted Internet resources. Censorship circumvention systems aim to defeat censorship by exploiting flaws in the network monitoring and interference systems deployed by the censors.
The most common techniques used by circumvention systems include tunneling, mimicry, and randomization 
(as briefly overviewed in Appendix~\ref{sec:circumvention_approaches}).

\paragraphb{Current circumvention development is \emph{reactive}:}
The gold standard of determining whether a censorship circumvention strategy successfully evades censorship in the real world is \emph{censorship measurement}~\cite{ooni, censoredplanet-10.1145/3372297.3417883, iclab-9152784, augur-7958591}, the process of executing various tests to gain information about a censorship apparatus.

The feedback gained from measurement is crucial to the development of circumvention strategies. However, over-reliance on this feedback results in a development process, where advancements in circumvention are \emph{reactive} to advancements in censorship.

We argue that such a reactive approach has given the censors the upper hand in the censorship arms race, as over-reliance on weaknesses of deployed censors leaves users vulnerable to sudden changes in censor strategy. This scenario can be seen in China, where censors suddenly deploy novel, advanced censorship strategies during sensitive political periods~\cite{Wu2023a}. 
We propound that \emph{the circumvention community should adopt proactive approaches to the design, analysis, and development of circumvention technologies} in order to complement\textemdash but not to replace\textemdash  existing reactive techniques such as censorship measurement.

In this work, our aim is to lay the foundations for proactive censorship research by building a generic framework for censorship emulation.

\paragraphb{\emph{Proactive} circumvention development with \name:}
We design and implement \name, a  \emph{generic censorship emulation framework} which is capable of emulating past, present, and futuristic (i.e., not-yet-deployed) censors.
\name enables robust testing of censorship circumvention and measurement techniques, especially techniques involving large traffic volumes such as protocol fuzzing or model extraction. 
Additionally, \name allows researchers to efficiently and safely test new anti-censorship techniques without incurring the various risks and pitfalls associated with real-world measurement. 
\name does not intend to replace real-world measurement and experimentation; rather, it aims to \emph{complement} the development of anti-censorship tools by (1) supporting the \textbf{emulation} of not just current censorship mechanisms, but also previously-deployed as well as not-yet-deployed (futuristic) censorship techniques; and 
(2) enabling the \textbf{testing} of circumvention and measurement techniques in a safe, isolated, and high-performance environment.
This aims to facilitate proactive anti-censorship research. 


\paragraphb{Overview of \name's architecture:}
\name is designed around a multi-layer architecture to provide an intuitive interface to the various layers at which censorship can occur. \name supports accelerated identifier-based filtering and censorship over Ethernet, IPv4, IPv6, TCP, and UDP identifiers. Furthermore, \name supports advanced censorship at the transport layer, which features two programmable interfaces.
Performance is critical when emulating censorship, as the overhead of network packet processing for an in-path censor can introduce subtle timing differences and potentially induce biased experiments. Therefore, we implement \name in Rust~\cite{rust}, a high performance systems programming language. 

\name enables developers and researchers to program arbitrary censorship behaviors through what we call \emph{censor programs} (Section~\ref{sec:censor_program}). 
We have implemented a high-level (based on Python) and a low-level (based on assembly language) programmable interface for censor programs. 
The first interface is build using RustPython~\cite{rustp}, a subset of Python. The Python API provides a high-level, user-friendly, moderately performant interface to network censorship programming. We refer to this interface as \textbf{PyCL}. PyCL allows developers and researchers to quickly and ergonomically test detection algorithms, especially when sensitive characteristics such as protocol timings are not a major concern.
%
Our second interface, \textbf{CensorLang}, is a custom assembly-like low-level interface. CensorLang is designed around two concerns: performance and static analysis. Although not as ergonomic as the high-level PyCL interface, CensorLang is much more performant, enabling experiments on a larger scale. Additionally, CensorLang is designed around the idea that censors value minimal state and minimal overhead when deploying censorship techniques. The design of CensorLang allows analysts to easily quantify the required execution time and state overhead of a censorship technique, which aids in the design of circumvention protocols. 

To provide researchers with an easy scripting interface to perform experiments, 
we have implemented \textbf{two IPC interfaces} (command-line and Python) that allow
 control and inspection of \name's internals at runtime. Additionally, we provide scripts designed to run isolated, reproducible experiments using \name.

In addition to standard censorship techniques, \name can 
emulate advanced data-driven (e.g., ML-based) censorship scenarios
by using  the ONNX~\cite{bai2019} library. 
One can  train models externally using libraries such as Tensorflow~\cite{tensorflow2015-whitepaper}, PyTorch~\cite{neurips2019-9015}, and scikit-learn~\cite{scikit-learn}, all of which support exporting models in the ONNX format. \name's ML support integrates seamlessly with censor programs by allowing programs written in both PyCL and CensorLang to execute models and make decisions based on their output.

In summary, \name aims to satisfy the following main \textbf{goals}: 
(1) support the emulation of all censorship mechanisms previously or currently deployed by real-world censors; 
(2) support emulating futuristic (i.e., not-yet-deployed) censorship strategies, in particular advanced data-driven censorship scenarios (such as flow correlation~\cite{nasr2018deepcorr, oh2022deepcoffea}, website fingerprinting~\cite{Attarian-Abdi-Hashemi-2019,Rimmer-Preuveneers-Juarez-Goethem-Joosen-2018,Sirinam-Imani-Juarez-Wright-2018,wangEffectiveAttacksProvable,hayes-danezis,Wang-Goldberg-2016}, and other ML-based traffic classifiers); (3) present an easy-to-use API for experimentation and extension by researchers and practitioners; and (4) be efficient and low-overhead.

 

\noindent \textbf{Showcasing \name:}
We demonstrate the capabilities  of \name  via a wide variety of censorship analysis scenarios encompassing both existing and futuristic censorship techniques.
First, we show how \name supports traditional censorship techniques by demonstrating its ability to perform HTTP keyword filtering, a common technique known to be used by the GFW~\cite{winter2012great}. We show that \name can perform this task efficiently, with 83\% fewer lines of code (LOC) and 20\% less overhead compared to Zeek~\cite{bro}, a general-purpose programmable DPI utility.

We also demonstrate \name's ability to emulate advanced DPI-based censorship techniques~\cite{10.1145/3419394.3423644, Wu2023a}. In particular, we show how the techniques recently used by the GFW in blocking fully encrypted protocols~\cite{Wu2023a} can be coded using \name's censor programs. As we show,  \name can emulate these with 58\% fewer LOC and 62\% less overhead than Zeek.
Note that while GFW stopped deploying these techniques recently, \name allows developers to experiment with them in a lab setting. 


We also show how \name can emulate futuristic censorship techniques through censor programs as well as its ML interface.  
In particular, we demonstrate how \name can trivially import the complex Triplet Fingerprinting (TF)~\cite{triplet-fingerprinting} model, and use it to perform state-of-the-art censorship. Compared to a similar program written in Scapy~\cite{rohith2018scapy}, \name performs TF-based censorship with 60\% fewer LOC and 10\% less overhead.

 
 \paragraphb{Paper's organization:} In Section~\ref{sec:motivation} we motivate the need to a censorship emulation framework like \name. 
 In Section~\ref{sec:design} we introduce the main components and capabilities of \name. 
 In Section~\ref{sec:censor_program} we introduce how Censor Programs are created in \name. 
 In Sections~\ref{sec:showcases},~\ref{sec:evaluations} we demonstrate \name's capabilities by performing a variety of censorship emulation experiments.
 We discuss ethics in Section~\ref{sec:ethics} and conclude in Section~\ref{sec:concl}.

\section{On The Need for a Censorship Emulation Framework}\label{sec:motivation}



We start by discussing shortcomings in the development and testing of censorship circumvention and measurement techniques. Next, we enumerate the features a censorship emulation framework must have to resolve these shortcomings. Finally, we discuss why existing efforts fail to satisfy these expected features.


\subsection{Shortcomings of Reactive Circumvention Development}\label{sec:reactive_challenges}


The community's reactive(-only) approach to censorship research lends itself to the following shortcomings.  


\paragraphb{Developed techniques may not be resilient:}
%
A censorship circumvention system is \textit{resilient} if it evades censorship across a wide variety of censorship scenarios. This is important for users of these tools, as loss of access can be difficult to restore without access to the free internet. One example of non-resilient censorship circumvention is that many popular censorship circumvention systems are vulnerable to statistical analysis~\cite{wang2015seeing}, likely because no existing censor has been observed using these techniques. However, as the capabilities of censorship systems improve, these attacks may become increasingly feasible, highlighting the need for circumvention systems to \emph{pre-emptively} develop countermeasures.

\paragraphb{Environmental Sensitivity:}
Censorship is constantly evolving, and the censor strategy, location, time, and vantage point used to measure censorship can drastically impact the results of circumvention and measurement experiments. In some cases~\cite{middleboxes-Raman2022a}, censors target residential ISPS, while cloud service hosting often~\cite{Wu2023a} used by researchers to perform measurements is not censored at all. Additionally, some censorship strategies are implemented intermittently, such as China's blocking of encrypted protocols, which is deployed both regionally and during politically sensitive events~\cite{Wu2023a}.

\paragraphb{In-the-wild measurement is costly:}
Whether using cloud services or cooperating with users in censored regions to install vantage points for measurement, maintaining access to a censored network for censorship measurement experiments can incur large costs. In general, because censorship measurement involves finding instances of censorship among a massive state space of network traffic, performing large-scale measurement requires sending and receiving huge quantities of network traffic. In scenarios where bandwidth is charged by usage, this can lead to very high costs. For example, a measurement system running on Tencent Cloud utilizing the full bandwidth of a 1Gbps link with no downtime could incur costs of over \$38,000 each month~\cite{tencent-cost}. While existing systems such as OONI~\cite{ooni-probe} may not utilize this level of bandwidth, more expensive measurement such as protocol-based measurement~\cite{10.1145/3419394.3423644} can easily use this much bandwidth. To put matters into perspective, a large-scale censorship measurement system should not only have vantage points in one country, but in every country around the world, or even every Autonomous System (AS). Overall, the costs of running a comprehensive censorship measurement platform are very high. The usage of remote vantage points, such as Quack's use of echo servers~\cite{Ben2018Quack} or remote measurement of DNS censorship does help to mitigate this, but not every AS has these servers available. 


\paragraphb{Safety risks to measurement parties:}
The most accurate way to measure censorship experienced by real people is to recruit volunteers living in censored areas to use their own personal internet connections and perform measurement on behalf of a measurement platform. This comes with personal risk to the volunteers~\cite{ethical-measurement}, as an ISP that observes a large and diverse number of forbidden connections for the volunteer's IP might decide to terminate their internet service or even report them to repressive authorities. If a cloud provider in a censored region detects researchers performing censorship experiments, they may decide to retaliate by taking action against their account, or even actively interfering with the accuracy of the results. For example, a researcher operating probes from a rented cloud server in China may have their account terminated if they send too much suspicious traffic. Additionally, a censor that observes the researcher's traffic may use these observations to resolve flaws in their censorship method before a circumvention method can be widely disseminated.
Because each censored request increases the likelihood a censor will pay attention to a particular host, the volume of traffic sent from vantage points when performing measurements is directly related to its risk. As such, minimizing this traffic using advanced measurement techniques is imperative to minimize risk to researchers and volunteers.

Testing and developing measurement techniques with maximum information, minimal runtime, and minimal risk requires a realistic censorship emulation framework that provides the same feedback as a real-world censor.

\subsection{Features Expected In An Emulation Framework}\label{sec:expected_features}



In order for a censorship emulation framework to meet the needs of circumvention and measurement researchers developing new techniques, it must meet the following conditions:

First, it must be \textbf{general}, which means it should support a wide variety of censorship techniques. Censorship is composed of two parts: the decision function and the action. One example of a decision function is whether a forbidden hostname is present in the SNI of a TLS \texttt{ClientHello} record. One action a censor could take in response to this decision function being triggered is to simply drop the packet, effectively terminating the connection. The decision functions  used in real-world censors can be simple, such as the detection of a forbidden IP~\cite{Verkamp2012a}, Host~\cite{Verkamp2012a}, or SNI~\cite{chai2019a}, but can also use more advanced techniques such as the presence of a forbidden phrase~\cite{winter2012great} or even a custom entropy function, as used in the GFW~\cite{Wu2023a}. \name supports identifier-based blocking from the IP level to the HTTP/DNS/TLS level, but is also capable of emulating advanced protocol-based and content-based decision functions via \emph{censor programs}, which we describe in Section~\ref{sec:censor_program}.

A censorship emulator should also be \textbf{extensible}, allowing emulation of techniques that are not yet deployed in real world censors. The efficiency/cost ratio of computing resources has increased dramatically over time~\cite{schaller1997moore}, and censorship strategies that are currently infeasible due to overhead may one day become commonplace. Website fingerprinting, an attack that infers a user's web browsing traffic despite encryption is often modeled as a small-scale or targeted attack due to the overhead of implementing it on network hardware, but if such an attack were to become efficient and accurate enough to be deployed at a large scale, it would have significant repercussions for both censorship and privacy. \name supports these futuristic scenarios with both \emph{censor programs} and its support for arbitrary ML models, described in Section~\ref{sec:censor_program_design}.

Additionally, a censorship emulator must be \textbf{easy to use} and provide a simple interface to describe censorship strategies. In order to promote a proactive approach to the development of circumvention tools, the barrier to development should be very low. Existing systems equipped for large-scale DPI such as Zeek~\cite{bro} often overly complicate or overwhelm users. We describe our efforts towards usability in Section~\ref{sec:python}.

Finally, a censorship emulator should be both \textbf{performant} and allow quantifying the overhead of a given censorship strategy. If operating over packet data and making decisions causes delays, packets could be retransmitted, changing the semantics of the underlying traffic. In sensitive experiments such as timing-based obfuscation protocols, this could drastically change traffic characteristics. \name's performance allows it to be used as a small-scale network appliance, processing traffic for an entire network. Additionally, our domain-specific language, CensorLang, is designed around static analysis, allowing researchers to easily quantify the overhead of a given censorship strategy. We describe our efforts towards performance and static analysis in Section~\ref{sec:censorlang}.

\subsection{Previous Attempts at Emulating Censorship}\label{sec:existing_efforts}


Here, we survey existing tools that may be used for censorship emulation, and discuss how they fall short in achieving the goals stated in Section~\ref{sec:expected_features} for censorship emulation (this is summarized in Table~\ref{tab:dpi_comparison}). 



\begin{table}[th]
    \centering
    \caption{Comparison of existing tools for censorship emulation.}
    \begin{tabularx}{\columnwidth}{|X|c|c|c|c|}
        \toprule
        \small{Tool} & \small{General} & \small{ML} & \small{Usable} & \small{Performant}\\
        \midrule
        Jafar & \xmark & \xmark & \xmark & \cmark \\
        Geneva Tests & \maybemark & \maybemark & \cmark & \xmark \\
        Zeek & \cmark & \xmark & \xmark & \cmark \\
        Scapy & \cmark & \maybemark & \cmark & \xmark \\
        MITMProxy & \xmark & \xmark & \cmark & \xmark \\
        \bottomrule
    \end{tabularx}
    \label{tab:dpi_comparison}
\end{table}
\paragraphb{Censorship-Dedicated Tools:}
We are aware of two prior works that perform limited censorship emulation. 
Jafar~\cite{jafar}, part of the OONI project~\cite{ooni}, is capable of locally simulating censorship actions, including drops based on IP or keyword, TCP resets based on IP or keyword, DNS hijacking, HTTP hijacking, and HTTPS hijacking. While these triggers and blocking actions represent the vast majority of actual censorship events, they do not flexibly represent all possible censorship strategies. While Jafar is fairly performant, it is primarily designed as a test suite of common and simple censorship methods, and is not designed to be extended to support advanced censorship scenarios.


Geneva~\cite{geneva-Bock2019a} uses genetic algorithms to evolve circumvention strategies against censors. It is capable of producing a wide variety of censorship circumvention strategies such as TCB Desync and TCB Teardown. However, Geneva includes a small suite of Python modules that emulate censors in order to test its capability to evolve known strategies. Each of these modules relies on two functions: \texttt{check\_censor} and \texttt{censor}. The first function, \texttt{check\_censor}, checks whether a connection should trigger a censorship action, while \texttt{censor} carries out the censorship action. Overall, while Geneva's test suite is theoretically capable of acting as a censorship emulation platform, its simplistic binary \texttt{check\_censor} function and performance~\cite{scapy-performance} are not satisfactory. Geneva's test suite largely relies on Scapy, so it suffers many of the same issues.


\paragraphb{General Purpose Network  Analysis Tools:}
Tools designed for general-purpose DPI such as Scapy~\cite{rohith2018scapy}, Zeek~\cite{bro}, and MITMProxy~\cite{mitmproxy} can also be used as platforms for censorship emulation. However, while general-purpose DPI systems often contain the raw capability to emulate censor behavior, they lack key features required to be useful as fully capable censorship emulation platforms.

Scapy~\cite{rohith2018scapy} is a Python library based around DPI. It supports monitoring, crafting, and injecting packets. Scapy provides a wide variety of packet dissectors, which provide high-level interfaces to various protocols such as TCP, TLS, and DNS. A common pitfall when using Scapy is that it suffers from fairly poor performance~\cite{scapy-performance}, which may introduce timing biases or even retransmissions if used as an on-path censor. The Scapy documentation offers some remedies for this issue, but claims: ``Scapy is not designed to be blazing fast, but rather easily hackable \& extensible. The packet model makes it VERY easy to create new layers, compared to pretty much all other alternatives, but comes with a performance cost.''~\cite{scapy-performance}.

Zeek~\cite{bro} is an open-source IDS featuring a powerful, custom scripting system that supports DPI at all layers from Ethernet to numerous application-layer protocols. It can issue actions such as blocking a connection or dropping all traffic from an IP address via its NetControl framework~\cite{netcontrol}. The NetControl framework is essentially a high-level interface that allows Zeek scripts to generically interface with firewalls and networking hardware (backends) using both high-level and low-level primitives. The full breadth of NetControl's ability to manipulate network traffic depends on which backend is used: some backends may not support certain features. Additionally, Zeek is a complex system with numerous processing layers, which significantly complicates the process of statically analyzing the overhead of a particular censorship strategy. Zeek does not support the use of ML, except via inefficient socket operations, so it remains incapable of emulating futuristic censors such those described by Wang et al.~\cite{wang2015seeing}.

MITMProxy~\cite{mitmproxy} is a proxy that intercepts HTTP and TLS traffic. It supports a Python scripting interface, and can perform censorship by terminating a connection, but the capabilities available to this interface are fairly minimal and isolated to HTTP and TLS specifically. This allows it to potentially analyze TLS-based protocols such as V2Ray~\cite{v2ray} or Tor~\cite{tor-269582}, but not others such as Obfs4~\cite{scramblesuit-10.1145/2517840.2517856} or Shadowsocks~\cite{shadowsocks}.

P4~\cite{bosshart2014P4} is a high-level programming language for packet processing. It is intended as a high-level interface that compiles to a variety of lower-level targets such as OpenFlow~\cite{mckeown2008OpenFlowEnablingInnovation}. Similarly to \name, P4 uses a match-action paradigm. While P4 is highly performant, it lacks support for ML classification.

In summary, while these general-purpose network interception tools work well for DPI, they are not well-suited to performing censorship emulation and experimentation. In the case of Scapy, this is largely due to performance problems. Zeek is fully featured and relatively performant, but suffers from a level of complexity that hinders analysis by ordinary practitioners/researchers and does not support emerging  data-driven censorship techniques. Finally, MITMProxy has limited protocol support and very limited ability to emulate known censorship strategies.

\section{\name: Components and Capabilities}\label{sec:design}
In this section, we describe the overall architecture of \name and how its layered design enables it to emulate past, present, and future forms of internet censorship. We also describe \name's features and capabilities at each level.
\subsection{\name's Overall Architecture}

\begin{figure*}
    \centering
    \includegraphics[width=1.6\columnwidth]{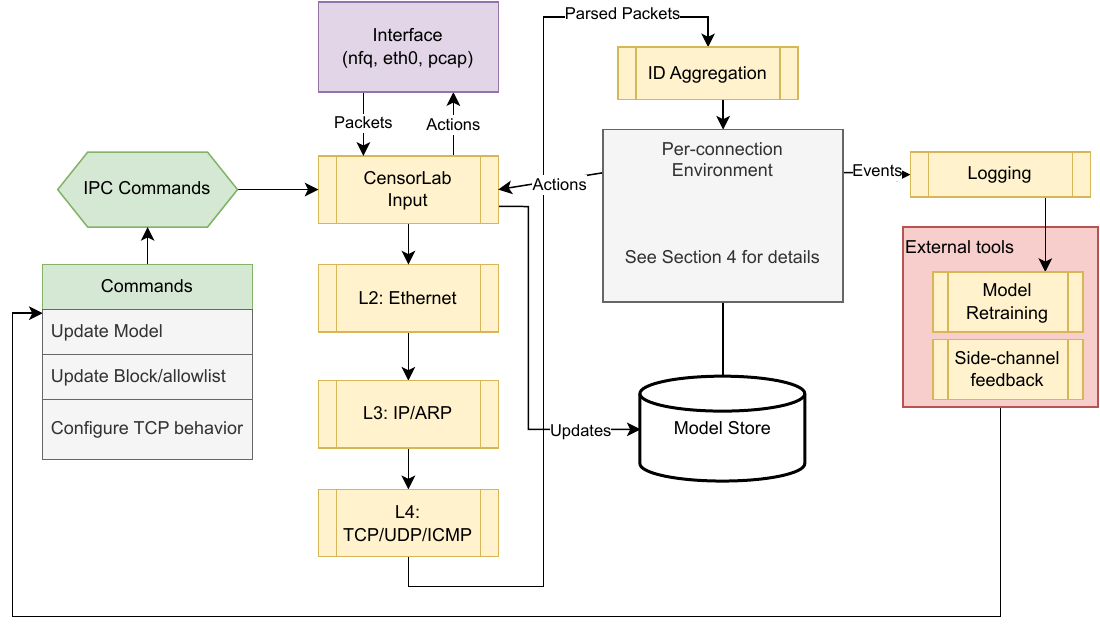}
    \caption{The overall architecture of \name. \name is split into 4 main components: the IPC interface, which controls \name at runtime, the per-layer packet processing, which handles parsing and identifier-based blocking, the per-connection environment, which performs the bulk of packet analysis, and the model store, which manages ML models. The details of the per-connection environment are explained further in Section~\ref{sec:censor_program}.}
    \label{fig:censorlab_architecture}
\end{figure*}

At a high level, \name acts as an arbiter of packets. Each packet processed by \name is subjected to parsing, block/allowlists, flow aggregation, per-connection processing, resulting in a per-packet {\tt action}. Packet processing in \name is done on the wire, meaning all processing must occur before any action is taken, including allowing the packet to be sent.

\begin{table}[ht]
    \centering
       \caption{\name supports automatic allow \& blocklisting on these identifiers. Layer 4 is split into three categories representing protocol, service, and connection blocking.}
    \label{tab:identifiers}
    \begin{tabular}{c|c}
        \toprule
        Layer & Identifier\\
        \midrule
        L2 Ethernet & MAC Address\\
        L3 IPv4/IPv6 & IP Address\\
        L4 TCP/UDP & Port\\
        L4 TCP/UDP & (IP, Port)\\
        L4 TCP/UDP & (IP, Port, IP, Port)\\
        \bottomrule
    \end{tabular}

\end{table}
Figure~\ref{fig:censorlab_architecture} plots the architecture of \name. Packets are introduced to \name via a generic packet ingestion interface, labeled \emph{Interface} in the diagram. Section~\ref{sec:modes} describes each of these interfaces in further detail. Upon receiving a packet from the interface, \name processes the packet layer-by-layer. Each layer is capable of emitting actions before the packet is processed by the next layer, in order to allow early identifier-based blocking. Each identifier is associated with both an allowlist and a blocklist, and these lists are evaluated according to Algorithm~\ref{alg:allow_block} immediately as the layer is processed. Blocklists are evaluated before allowlists, and each has an action associated with rejection, denoted $Action_{block}$ and $Action_{not\_allowed}$, respectively. The identifiers supported by allowlists and blocklist at each layer are listed in Table~\ref{tab:identifiers}.

\begin{algorithm}
\caption{Allowlist + Blocklist Execution}\label{alg:allow_block}
{\small
\begin{algorithmic}
\If{$Identifier \in Blocklist$}
    \State \Return $Action_{block}$
\ElsIf{$Identifier \notin Allowlist$}
    \State \Return $Action_{not\_allowed}$
\EndIf
\end{algorithmic}
}
\end{algorithm}

The \textbf{Layer 2} component of \name handles Ethernet, the most common data link layer protocol. This layer processes the three fields most commonly contained in Ethernet traffic: source address, destination address, and EtherType~\cite{ieee-ethernet}. For simplicity, this layer does not process Ethernet frames that specify a length rather than an EtherType~\cite{ieee-ethernet}, nor does it process 802.1Q frames~\cite{ieee-8021q}. The source and destination addresses are not likely to be used in censorship, but still have allow/blocklists, as they may be useful for isolating traffic while performing experiments. After processing the identifiers, the payload of the Ethernet frame is forwarded to a Layer 3 protocol handler based on the EtherType field.

\begin{table}[ht]
    \caption{Layer 3 Features}
    \label{tab:l3_features}
    \centering
    {\small
    \begin{tabular}{|c c c|}
        \toprule
        IPv4 & Both & IPv6 \\
        \midrule
        DSCP & Version & Traffic Class\\
        ECN & Header Length & Flow Label\\
        IPID & Total Length &\\
        DontFrag & TTL &\\
        MoreFrags & Next Header ID&\\
        Checksum & Source IP &\\
         & Dest IP &\\
         & Payload &\\
        \bottomrule
    \end{tabular}
    }
\end{table}

The \textbf{Layer 3} component of \name handles IPv4, IPv6, and ARP traffic. Table~\ref{tab:l3_features} lists the features extracted by this component. In particular, the source and destination addresses are considered identifiers. In addition, Layer 3 also allows the use of subnets as identifers. If the Layer 3 protocol is IPv4 or IPv6, the ``Next Header ID'' field is used to identify the contained protocol, and pass the payload to the Layer 4 component. Due to its local nature and irrelevance to internet censorship, ARP is parsed for validity but otherwise ignored by \name.

\begin{algorithm}
\caption{Layer 4 Packet Aggregation Scheme}\label{alg:l4agg}
{\small
\begin{algorithmic}
\State $IP_{src}, IP_{dst} \gets L3.IP_{src}, L3.IP_{dst} $ \Comment{Generic to IPv4/6}
\State $Port_{src}, Port_{dst} \gets L4.{src}, L4.{dst}$ \Comment{Generic to TCP/UDP}
\If{$Port_{src} < Port_{dst}$}
    \State $IP_1 , IP_2 \gets IP_{src}, IP_{dst}$
    \State $port_1 , port_2 \gets port_{src}, port_{dst}$
\Else
    \State $IP_1 , IP_2 \gets IP_{dst}, IP_{src}$
    \State $port_1 , port_2 \gets port_{dst}, port_{src}$
\EndIf
\State $Key \gets (IP_1, IP_2, port_1, port_2, L3.NextProto)$
\If{$Key \notin Connections$}
\State $Connections[Key].packets \gets []$
\State $Connections[Key].port_{src\_initial} \gets port_{src}$
\EndIf
\end{algorithmic}
}
\end{algorithm}

The majority of \name's functionality occurs in the \textbf{Layer 4} component, which handles transport-layer protocols: TCP and UDP. In this layer, packets are aggregated into ``connections'' based on a unique identifier constructed from the packet's source IP, destination IP, source port, destination port, and transport protocol. This identifier is constructed based on Algorithm~\ref{alg:l4agg}. By using the sorted IP and port pairs, all packets belonging to the same connection are classified as the same connection regardless of the packet's direction. When the first packet associated with a connection is processed, its source port is also stored. This stored port can later be used to determine whether any future packet in the connection is being sent from the client that initiated the connection (initiator) or from the other side of the connection (responder), terms also used by Zeek~\cite{bro}. For client-server protocols such as HTTP, TLS, or DNS, the role of initiator and responder is identical to the notion of client or server, respectively. 

Before connection-level processing, the packet is checked against the Layer 4 allow and blocklists. These lists work on three levels: port, service, and connection, corresponding to the last three entries of Table~\ref{tab:identifiers}. Port-level blocking targets all traffic to and from a specific port. This is useful when blocking the default port of a protocol entirely, or for ignoring or dropping noisy broadcast traffic in a live environment. Service-level blocks target a specific port on a specific host. This type of block is often~\cite{10.1145/3419394.3423644} used when a censor detects that a particular server is hosting a circumvention service. The connection level blocklist targets a specific instance of a connection, and are typically activated in a dynamic, rather than static manner. Once a connection is deemed forbidden, all future packets of that connection will be acted upon in some way by the censor. Unlike the other identifiers, there is no connection-level allowlist, as the exact details of a TCP or UDP connection are rarely known ahead of time.

\paragraphb{Censor Program:} Censor programs in \name allow users to programmatically define the per-connection behavior of the censor. When the aggregation algorithm has identified a new connection, a \emph{per-connection environment} is created for it. After the per-connection environment is initialized, each packet recognized as part of that connection is used as the input of the censor program, provided the packet has not been filtered by other means such as a blocklist.  This program executes once per packet, and maintains its state across packets. The details of censor programs are elaborated further in Section~\ref{sec:censor_program}. Each execution of the censor program outputs an action, with ``Allow'' being the default action.

\begin{table}[!t]
\centering
\caption{Supported actions in \name}\label{tbl:process_return_values}
{\small
\begin{tabular}{c|c}
    \toprule
    Value & Action\\
    \midrule
    None & No action taken\\
    ACCEPT & Forward the packet\\
    DROP & Drop the packet\\
    RESET(N) & Send N TCP RST packets to both ends of the\\
             & connection based on the current packet\\
    DELAY(N) & Wait N seconds before forwarding this packet.\\
    \bottomrule
\end{tabular}
}
\end{table}

\paragraphb{Supported actions:} Finally, once the program has been run, and an action is determined, this action will be executed. The list of supported actions is shown in Table~\ref{tbl:process_return_values}

Drop and Reset in particular are based on observation of real-world censorship. Dropping packets is often used in the Great Firewall (GFW) for censoring encrypted connections to censored services or circumvention proxies, while sending TCP RST packets is used in the GFW for unencrypted HTTP traffic where forbidden keywords are detected~\cite{winter2012great}. The Delay action has seen more use in recent years, e.g., in Russia where traffic to Twitter was throttled rather than blocked~\cite{xue2021throttling}.

In addition to the main set of actions, a program may also inject raw ethernet packets. This may be done an arbitrary number of times for each program, and is supported only in PyCL. 

\subsection{\name's Modes of Operation}\label{sec:modes}

\begin{figure}[t!]
    \centering
    \includegraphics[width=0.70\columnwidth]{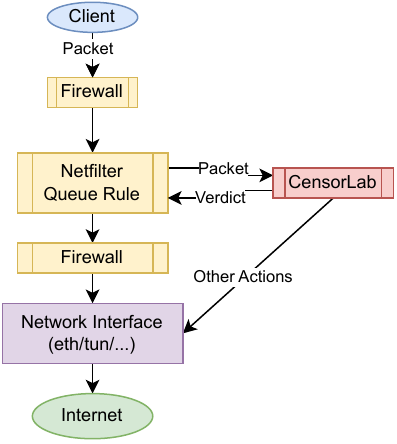}
    \caption{In \texttt{Tap} mode, \name intercepts packets using a netfilter queue firewall rule, and issues verdicts for each packet. \name is also capable of interfacing with the output interface directly for scenarios not supported by netfilter queues, such as injecting multiple packets.}
    \label{fig:censorlab_arch_nfq}
\end{figure}
\begin{figure}[t!]
    \centering
    \includegraphics[width=0.70\columnwidth]{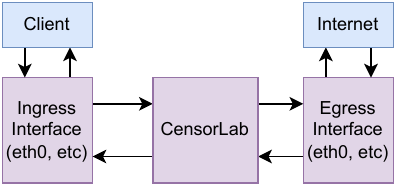}
    \caption{In \texttt{Wire} mode, \name acts as an intermediary between two physical network interfaces. Packets received from the ingress interface, if accepted, are written to the egress interface and vice versa.}
    \label{fig:censorlab_arch_wire}
\end{figure}

\name has three modes of operation, \emph{Tap}, \emph{Wire}, and \emph{PCAP}. 
For the majority of censorship experiments, the most suitable \name mode is \textbf{Tap mode}, which uses netfilter queues (NFQs) to interact with network traffic. Netfilter queues receive traffic by configuring \texttt{iptables} or \texttt{nftables} to forward traffic to the queue. NFQs act as a userspace firewall, where each packet is sent to a userspace program to be processed. Actions such as forwarding, modifying, and dropping packets are issued via \textit{verdicts}~\cite{iptables}. However, not all actions supported by \name are also supported as NFQ verdicts, and so \name must also be given direct access to a network interface.

The overall architecture of \name in Tap mode is shown in Figure~\ref{fig:censorlab_arch_nfq}. In \textit{Tap} mode, the \texttt{Accept} action triggers an \texttt{Accept} verdict, while the \texttt{Drop} action issues a \texttt{Drop} verdict. Because NFQ is designed around issuing a single verdict per packet, the \texttt{Reset} action must perform out-of-band activities. First, \name will issue a \texttt{Drop} verdict for the packet. Next, it will use information made available in the Linux routing and ARP table to identify how the TCP RST packets must be routed.  RST packets are then injected directly onto the given interface using the raw socket API. Finally, the \texttt{Delay} action performs a \texttt{Drop} verdict, but copies the packet's data before sending it to a background thread, which schedules packets based on the given delay. Delay allows for complex timing-based behaviors, though traditional throttling techniques may be implemented for loss-tolerant protocols (e.g. those using TCP or an application-layer loss detection over UDP) using the \texttt{Drop} action.

Though Tap mode can be configured to operate on networking hardware such as a router, it is generally intended for use on a single system. Tap mode can be used to route all traffic through it in order to test the resilience of a circumvention tool, or even in an isolated scenario such as Docker, where most or all network interfaces are virtualized. Tap mode offers the most flexibility to the user, as they can choose which traffic is forwarded through \name via iptables or nftables rules. By default, the scripts provided with \name will result in all system traffic being forwarded through \name.

For experiments that require a more realistic censorship scenario, \name can act as a network appliance in \textbf{Wire mode}. 
In Wire mode, \name forwards packets between two different interfaces, acting as an active ``man in the middle'' eavesdropper between them. In Wire mode, the \texttt{Accept} action result in a packet being written to the other interface normally, while the \texttt{Drop} action results in no action at all, as the packet is not forwarded. The \texttt{Reset} action injects TCP RST packets on both interfaces, while the \texttt{Delay} action sends packets to a scheduling thread, where they are forwarded to the other interface after the given amount of time has elapsed.

The overall architecture of \name in Wire mode is shown in Figure~\ref{fig:censorlab_arch_wire}, where \name is shown to act as an arbiter between the two interfaces. In this diagram, we distinguish between the two interfaces as ``ingress'' and ``egress''. While \name contains no handling specific to either interface, the direction of traffic is calculated and passed to further layers.

Finally, \name also supports a \textbf{PCAP mode}, where instead of physical interfaces, it reads a PCAP or PCAPNG file, and processes each packet as if it were operating on a real network. Because a PCAP is a record of past traffic, \name cannot interfere, only log what it would have done in a given scenario. In PCAP mode, each negative action (\texttt{Drop}, \texttt{Reset}, \texttt{Delay}) is logged using \name's logging facilities, and can be used for later analysis.
Because interfering actions may affect future network traffic in a realistic environment, care must be taken when interpreting the output of this mode. For example, PCAP mode can be used to test the efficacy of a passive network protocol classsifier, but cannot be used to test whether dropping packets triggers a circumvention protocol to exhibit specific behaviors.
By default, \name operates opportunistically in PCAP mode, and will process all packets in the file immediately, but ``time emulation'' is also supported. When using time emulation, \name will process packets at times relative to the recorded timestamps in the PCAP file. Additionally, when operating in time emulation mode, \texttt{Delay} will be simulated properly. When operating in opportunistic mode, \name will emit a log message both when $P$ is processed and immediately before processing $P'$ where $P'$ is the earliest packet such that $P'\times timestamp > P\times timestamp + delay$.

\subsection{Allow \& blocklists}
Allow \& blocklists enable \name to perform terminating actions based on identifiers present in Ethernet, IPv4, IPv6, TCP, and UDP headers. Each blocklist has a corresponding action: $Action_{block}$, which specifies the action to be taken if the given identifier is on the blocklist. Each allowlist has a corresponding $Action_{not\_allowed}$ which specifies the action to be taken if an identifier is not on the allowlist. Each of these lists can be modified in three ways: config, IPC, and censor program. Allow \& blocklist entries and actions in the \name config file: \texttt{censor.toml} are loaded when the program starts. These lists can also be modified using the IPC interface, as specified in Appendix~\ref{appendix:ipc_commands}. Finally, the allow \& blocklists can be manipulated using censor programs, using special functions and commands specified in Appendix~\ref{appendix:ipc_commands}.

\subsection{\name's IPC Interface}\label{sec:ipc}
\name can be controlled externally using the IPC interface, which communicates over TCP on \texttt{127.0.0.1}, using port 25716 by default. The loopback interface was chosen in order to minimize interference with network traffic operation, and for security reasons. Because \name is technically capable of intercepting traffic on loopback interfaces, it is configured to ignore its own IPC traffic by adding a service-level blocklist entry. In the case that a user performs censorship experiments with \name using the loopback interface, they must take care to avoid potential port collision, or use alternative loopback addresses such as \texttt{127.0.0.2}. We provide two client interfaces to the \name IPC interface: command-line (CLI) and Python. The CLI interface is a program and can be invoked from an interactive shell or shell script, while the Python interface is intended to be used in more complex scripts, such as those coordinating large-scale experiments. The IPC interface is fully specified in Appendix~\ref{appendix:ipc_commands}. In summary, it can be used to terminate \name, trigger a configuration reload, manipulate allow \& blocklists, load programs, debug programs, and manipulate the model store.

\section{Censor Programs}\label{sec:censor_program}

In \name, censor programs act as a 
programmable  interface to allows researchers to  model  censorship scenarios. The design of censor programs is centered around satisfying the following goals:



\subsection{Goal 1: Capablity to Model a Wide Variety of Censor Behaviors}\label{sec:censor_program_design}
The primary goal of \name is to provide a comprehensive censorship emulation platform. In this section, we describe the execution environment of censor programs, and how censor programs satisfy this goal. The over-arching architecture of the censor program execution environment is generally identical across the PyCL and CensorLang interfaces, though we will elaborate any differences.

\begin{figure}[!ht]
\centering
\begin{subfigure}[b]{0.4\columnwidth}
\begin{minted}[fontsize=\small]{python}
n = 0
def process(packet)
  if n >= 10:
    return DROP
  n += 1
\end{minted}
\caption{PyCL}\label{fig:simple_censor_program}
\end{subfigure}
\hfill
\begin{subfigure}[b]{0.45\columnwidth}
\begin{minted}[fontsize=\small]{nasm}
COPY reg:u32:0 0
process:
if reg:u32:0 GEQ 10:
    return DROP
INC reg:u32:0 
\end{minted}
\caption{CensorLang}\label{fig:simple_censor_program_censorlang}
\end{subfigure}
\caption{Censor program to drop connections longer than 10 packets}
\end{figure}

From a censor's point of view, the ideal censorship algorithm is able to terminate forbidden connections based on a single packet, without connection tracking. This is a per-packet censor, which can be represented as simply $f(Packet) \rightarrow Action$. One example of this is WireGuard VPN traffic~\cite{donenfeld2017wireguard}, where the protocol may be reliably detected by checking the UDP payload for the pattern \begin{verbatim}^[\x01-\x04]\x00\x00\end{verbatim}. A more complicated instance of this is the GFW censorship of encrypted tunneling protocols such as ShadowSocks~\cite{Wu2023a}, which uses a series of heuristics over the metadata of a single data packet. However, many traffic analysis algorithms operate on a per-connection basis, and require metadata from \emph{multiple packets}~\cite{wang2015seeing}. To support this, a censor must maintain some notion of per-connection \emph{state}, an accumulating set of variables that may be modified for every processed packet. This can be modeled as $f(State, Packet) \rightarrow (State, Action)$, where each packet is part of the same connection. The most common instance of state in DPI systems is the TCP Control Block or TCB, which maintains the state required to interpret the current status of a TCP connection~\cite{geneva-Bock2019a}. Disrupting the TCB has been used before as a censorship circumvention strategy~\cite{geneva-Bock2019a}. An example of this is detecting forbidden keywords in TCP protocols such as HTTP. In order to accurately block traffic that mentions forbidden keywords, a censor must maintain an accurate TCB, as the forbidden text may span across multiple packets. Some traffic analysis attacks expand beyond per-connection analysis, such as a flow correlation attacks against anonymity networks~\cite{nasr2018deepcorr,oh2022deepcoffea} or statistical analysis of the traffic sent to potential proxies. These attacks can be modeled the same way as per-connection analysis, except they share state across all intercepted packets rather than only packets from the same connection. For simplicity, performance, and because per-connection analysis encompasses a wide variety of use cases, we have chosen to limit programmable censor state to a per-connection scope. 

We define censor programs in terns of per-connection state by splitting their execution into two components: initialization and per-packet processing. This is similar to programming paradigms popular in embedded programming such as Arduino~\cite{arduino}, which uses an initialization-loop paradigm, or DPIs such as Zeek, which has an initialization function and a per-packet function among many other functions supported by its event system. Figure~\ref{fig:simple_censor_program} demonstrates a simple censor program written in PyCL. Rather than define multiple functions: one for initialization and one for packet processing, PyCL programs simply define the packet processing function. This is purely a stylistic choice to avoid the need for python's \texttt{global} qualifier. When a connection is first initialized, following the aggregation described in Algorithm~\ref{alg:l4agg}, a new PyCL execution environment is initialized, and the code snippet is executed in its entirety. For the first and all subsequent packets, the \texttt{process} function is called within that environment. Because each connection receives its own Python execution scope, state is scoped to connections. In addition to \textit{per-connection} state, \name also allows storing state on a per-host basis. This allows for decisions such as blocking hosts that receive traffic detected as proxy traffic too many times.
In this program, a single state variable \texttt{num\_packets} is defined and initialized to zero. As each packet is received, the variable is incremented, and once the variable reaches 10, all subsequent packets in the connection are dropped. For the sake of cleaner code, we treat the \texttt{None} return value as an \texttt{Accept} action; by default, the processing function will accept the packet. We describe the various actions in Table~\ref{tbl:process_return_values}. The notation described in this table for actions carries over to CensorLang programs as well. Figure~\ref{fig:simple_censor_program_censorlang} shows the same censor program as Figure~\ref{fig:simple_censor_program}, except written in CensorLang. In CensorLang, the \texttt{process} label is used to denote the start of the packet processing function, and all code before this label is treated as initialization code. CensorLang is meant to resemble simple, opcode-based programming languages such as assembly languages, and is described in more detail in Section~\ref{sec:censorlang}.



\subsection{Goal 2: Ease of Use}\label{sec:python}
Though \name is intended for rigorous experimentation involving censorship and anti-censorship systems, we also encourage its use as a development tool for censorship circumvention systems. To this end, programming \name should be simple, and easy to use. The programming paradigm used by censor programs is well-known for its ease of use, as seen by the prominence of similar paradigms in education~\cite{el2017review}. We further enhance the accessibility of censor programs by allowing the use of PyCL.

The PyCL implementation of censor programs is written using a subset of Python, specifically the subset supported by RustPython~\cite{rustp}, an embeddable Python interpreter. We chose RustPython for its ease of integration. In order to maximize the performance of censor programs written in PyCL, we take advantage of RustPython's compilation and JIT functionality. When a PyCL censor program is loaded, we use RustPython to compile it as a code object. That code object is then reused for each new per-connection scope, removing the need to recompile. In addition to compilation, RustPython supports JIT compilation of Python functions. This support is experimental: If successful, it leads to performance gains; otherwise the failure is logged, and the code executes normally.

\subsection{Goal 3: Statically Analyzable Overhead}\label{sec:censorlang}
Censorship and anti-censorship are a series of trade-offs. Censors weigh overhead, accuracy, and false positive rate (FPR), while circumvention systems weigh overhead, security, and evasion. To guide the development of circumvention systems, metrics such as overhead and false positive rate for a given censorship strategy should be easy to obtain.

While writing censor programs in PyCL is accessible and moderately performant, as shown in Section~\ref{sec:showcases}, its dynamic type system means that certain optimizations available to more rigorously typed languages cannot be applied. Additionally, its dynamic nature also makes static analysis of PyCL code much more difficult. In response to these issues, we introduce our simple, custom-built programming language designed specifically for censorship: \emph{CensorLang}. CensorLang is inspired by assembly language, with some inspiration taken from Rust's type system. The major choices behind CensorLang are as follows:
\begin{compactitem}
    \item State is stored in typed, indexed registers such that counting registers used by the program can immediately quantify the memory overhead of the program;
    \item CensorLang programs cannot loop, so the performance overhead of the program can be expressed as  the dot product of its operations and their overheads.
\end{compactitem}
Figure~\ref{fig:censor_program_grammar} in Appendix~\ref{appendix:censorlang_grammar} describes the grammar of CensorLang. We compare the performance of CensorLang, PyCL, and other DPI systems in Section~\ref{sec:showcases}.

When a censor designs a new censorship algorithm, it must optimize for two properties: \emph{efficacy} and \emph{performance}. First, the censorship algorithm must be \emph{effective}: it must maximally interfere with forbidden network traffic while minimally interfering with allowed traffic. This is often quantified using measures such as accuracy, false positive rate (FPR), or the area under the precision-recall curve (PR-AUC)~\cite{dixon2016network}. The false postive rate of a censorship system is considered particularly significant for censors, as even very small FPRs can cause massive disruptions at large scales~\cite{dixon2016network}. Second, the censorship algorithm must be \emph{performant}. When we consider censorship, it is often in the context of nation-states that operate massive censorship apparatuses, such as the GFW. These censorship apparatuses process massive amounts of network traffic, and any overhead can result in significant degradation in network performance. Additionally, the processing power required to operate DPI-based classification on a massive scale can be prohibitively expensive, or even impossible for a network operator that handles a sufficiently large volume of traffic. To counteract this, the designers of anti-censorship systems are tasked with two intersecting objectives: they must minimize the accuracy that a censorship algorithm can achieve while maximizing the processing power required to do so. In addition, they must balance other concerns such as client overhead, privacy, security, anonymity, and cost. The processing and memory overhead of a censorship algorithm can be measured either experimentally or through static analysis. The PyCL interface to censor programs is easy to use, and reasonably performant, but performing static analysis on PyCL code is complex due to its dynamic type system~\cite{gulabovska2019survey}.

\subsection{Goal 4: Support for ML Censorship}\label{sec:ml_goal}
ML has become increasingly prominent in traffic analysis, and statistical analysis has repeatedly proven to be a powerful tool for network protocol classification~\cite{wang2015seeing}. While there are no known usages of ML in real-world censorship systems, Moore's law~\cite{schaller1997moore}, economies of scale, and the recent popularity of ML implies that ML censorship may one day be feasible at nation-state scale. Thus, we believe a censorship emulation system must support the use of ML models.

The ML component of \name is built around ONNX~\cite{bai2019}, a portable format for representing and executing ML models, as well as our own custom format for specifying how features are loaded into the model. Most major ML systems~\cite{tensorflow2015-whitepaper, neurips2019-9015, scikit-learn} support ONNX as an export format, allowing \name to use virtually any model from a wide variety of environments. Models can be loaded into \name via the IPC interface or via the configuration file. For simplicity, ONNX models must accept a $1 \times N$ tensor of 32-bit floats as input, and output a $1 \times M$ tensor of 32-bit floats. As described in the IPC specification in Appendix~\ref{appendix:ipc_commands}, each model is given a name when loaded in order to allow the use of multiple models. The syntax to evaluate a model in PyCL is:
\vspace{-2mm}
\begin{minted}{python}
out = model("model_name", [x,y,z,w])
\end{minted}
\vspace{-2mm}
The first argument specifies the model name, while the second is a list of arguments, which MUST match the model's shape, or the censor program will terminate for the given packet, using the default \texttt{Accept} action. The output of this function is a list of floats, as output by the model.

Using models in CensorLang is more complicated, owing to its register system. CensorLang provides a set of input and output registers for each model. For example, to execute a model \texttt{wf} that accepts three values and outputs two values, the syntax is:
\vspace{-2mm}
\begin{minted}[fontsize=\small]{nasm}
COPY 10 model:wf:in:01
COPY 0 model:wf:in:10
COPY 20 model:wf:in:22
MODEL wf
COPY model:wf:out:0 reg:f32:0
COPY model:wf:out:1 reg:f32:0
\end{minted}
\vspace{-2mm}

When a CensorLang program is loaded, \name is able to determine which models are used, and so per-connection state registers are only allocated as required.
\section{Showcasing \name's Performance}\label{sec:showcases}
In this section, we demonstrate the usability and extensiveness of \name by demonstrating its use across a wide range of censorship scenarios. We evaluate and compare \name using both the PyCL and CensorLang variants against both Scapy and Zeek in terms of \emph{performance} (time taken to process a pcap file), \emph{complexity} (lines of code), and qualitative aspects regarding \emph{usability}.


\begin{table*}[t]
    \centering
    \caption{Performance comparisons between \name and other DPI programs. Each scenario is reported in terms of time elapsed, in seconds. N/A indicates a lack of support for the required features for a given experiment.}
    {\small
    \begin{tabularx}{.94\linewidth}{c|p{3cm}|p{3cm}|p{3cm}|p{3cm}}
    \toprule
             & \multicolumn{2}{c|}{\name} &      &      \\
    Scenario & PyCL & CensorLang & Zeek & Scapy\\
    \midrule
    SNI Filtering & 389.65 & 301.92 & 450.60 & 480.76 \\
    Shadowsocks & 320.20 & 258.91 & 635.00 & 789.2 \\
    DNS Injection & 32.78 & N/A & N/A & 45.96\\
    ML Protocol Classification & 596.13 & 512.20 & N/A & 650.64\\
    \bottomrule
    \end{tabularx}
    }
    \label{tab:performance_comparison}
\end{table*}
\begin{table*}[t]
    \centering
    \caption{Lines of code for each of the programs listed in Table~\ref{tab:performance_comparison}.}
    {\small
    \begin{tabularx}{.94\linewidth}{c|p{3cm}|p{3cm}|p{3cm}|p{3cm}}
    \toprule
             & \multicolumn{2}{c|}{\name} &      &      \\
    Scenario & PyCL & CensorLang & Zeek & Scapy\\
    \midrule
    SNI Filtering & 6 & 7 & 12 & 26 \\
    Shadowsocks & 26 & 46 & 72 & 48 \\
    DNS Injection & 10 & N/A & N/A & 30\\
    ML Protocol Classification & 30 & 53 & N/A & 75\\
    \bottomrule
    \end{tabularx}
    }
    \label{tab:loc_comparison}
\end{table*}

\subsection{Keyword Filtering: TLS SNI}\label{sec:keyword}
Keyword filtering is a classical mechanism for Internet censorship. For example, the GFW is known~\cite{winter2012great}  to use keyword filtering  on raw HTTP traffic, performing censorship when forbidden terms are detected. As HTTPS becomes ubiquitous, censors perform filtering on other plaintext metadata, such as the SNI field of TLS traffic~\cite{chai2019a}. In \name, keyword filtering may be represented as a censor program utilizing regular expressions. A program that performs SNI filtering, approximating the GFW behavior described in~\cite{chai2019a} is:

\begin{minted}{python}
r = Regex(
  "^[\x16\x17]\x03[\x00-\x09]" +
  "google.com|facebook.com|..."
)

def process(packet: Packet):
  if 443 not in (tcp.src, tcp.dst):
    return ALLOW
  if r.matches(packet.tcp.payload):
    return RESET
\end{minted}

The regular expression used in this program utilizes a known, minimal TLS signature detection strategy used by the GFW, as shown in~\cite{Wu2023a}, followed by a plaintext check for forbidden domains. This program is not capable of performing filtering on TLS traffic utilizing the ESNI feature. In this experiment, we first extract a list of 1000 hostnames from the Tranco~\cite{tranco-lepochat2019} list of popular websites capable of responding to HTTPS requests without error.
We then use TCPDump~\cite{tcpdump} to capture all traffic utilizing port 443, and issue HTTPS requests to each domain using Firefox automated by Selenium~\cite{selenium}. For this experiment, we configure Firefox to disable the use of ESNI. To add complexity, these requests are performed in parallel using 8 Firefox tabs simultaneously. Finally, we use \name in PCAP mode, Zeek~\cite{bro}, and a program written using Scapy to perform the HTTP keyword fingerprinting task. The performance results of this experiment are shown in Table~\ref{tab:performance_comparison}, while the LOC for each program Table~\ref{tab:loc_comparison}. In qualitative terms, \name performs much of the DPI and packet capture work, allowing \name programs to be shorter and easier to implement.

\subsection{Response injection: DNS}\label{sec:dns_showcase}

In this section, we demonstrate how \name can be used to model more active forms of interference. In this case, we demonstrate how \name can be used to emulate DNS response injection. To begin, we define the following program, which injects \texttt{NXDOMAIN} responses for a small number of hosts. We benchmark the performance of PyCL, CensorLang, and Scapy on live traffic by performing 100000 DNS queries to DNS servers randomly chosen from \texttt{[8.8.8.8, 8.8.4.4, 1.1.1.1, 1.0.0.1, 9.9.9.9]}, using a multithreaded, asynchronous query program. These domains are chosen from the top 100000 entries in the Tranco list~\cite{tranco-lepochat2019}, with a randomly selected half of them designated as forbidden domains, and the other half designated as allowed domains. We exclude Zeek from this analysis, as we are not aware of functionality that would allow DNS response injection within Zeek directly. The following code is used as the PyCL censor, with other censors having the same functionality:

\begin{minted}{python}
r = Regex("google.com|facebook.com|...")

def process(packet: Packet):
    if udp.dst != 53:
        return
    match = r.match(packet.udp.payload)
    if match is not None:
        resp = packet.make_udp_response()
        resp.payload = dns.response(
            match, dns.NXDOMAIN
        )
\end{minted}

The regular expression's text is reduced for clarity, but is constructed using the list of forbidden domains. The results of this experiment are shown in Table~\ref{tab:performance_comparison} shows that PyCL is both more performant and concise compared to the equivalent Scapy program.

\subsection{Protocol Censorship: Fully Encrypted Traffic}\label{sec:protocol_censorship}

A recent study by Wu et al.~\cite{Wu2023a} discovered complex mechanisms used by the GFW to 
identify and block the fully encrypted, randomized protocols used by popular circumvention tools such Shadowsocks~\cite{shadowsocks} and Obfs4~\cite{scramblesuit-10.1145/2517840.2517856}.
These protocols fully encrypt traffic to minimize points of identifiability identifiers such as headers and handshakes, but the GFW is shown to have leveraged advanced mechanisms to identify these circumvention protocols~\cite{Wu2023a}. In this section, we demonstrate how \name can emulate GFW's censorship mechanisms discovered by Wu et al.~\cite{Wu2023a} to block popular circumvention protocols. Specifically, Wu et al. discover five exemption rules (Algorithm 1,~\cite{Wu2023a}) that allow non-circumvention traffic, and block the rest.

The first exemption rule checks the entropy of the first packet in a connection to be within the range $[3.4, 4.6]$. The third exemption rule is if the percentage of printable characters in the packet's payload is greater than 50\%. These exemption rules can be emulated in \name using the following censor program:  
\begin{minted}[fontsize=\small]{python}
r = Regex(
  "^([\x20-\x7E]{6} | .*[\x20-\x7E]{20}.* |" +
  "HTTP | [\x16\x17]\x03[\x00-\x09])"
)
def process(packet: Packet):
    popcount_upper = 4.6 * packet.payload.len
    if r.matches(packet.tcp.payload):
        return ALLOW_ALL
    printable, popcounts = 0,0
    for b in packet.payload:
        if 0x20 <= b <= 0x7E:
            printable += 2
            if printable > packet.payload.len:
                return ALLOW_ALL
        popcount += b.bit_count()
        if popcounts >= popcount_upper:
            return ALLOW
    if popcounts <= 3.4 * packet.payload.len:
        return ALLOW_ALL
    return DROP
\end{minted}

In the above regular expression, the expression \verb|[\x20-\x7E]{6}| matches any packet that begins with 6 printable ASCII characters (Exemption\#2, \cite{Wu2023a}), 
and the expression \verb|.*[\x20-\x7E]{20}.*| matches any contiguous sequence of 20 printable ASCII characters (exemption\#4, \cite{Wu2023a}).
Finally, \verb|HTTP| and \verb|[\x16\x17]\x03[\x00-\x09]| match HTTP and TLS packets respectively (exemption\#5, \cite{Wu2023a}).

We perform an experiment similar to the one performed in Section~\ref{sec:keyword} in order to evaluate \name using the same metrics. Using the same list of domains from Section~\ref{sec:keyword}, we use TCPDump~\cite{tcpdump} to monitor all outgoing traffic utilizing either port 443 or our Shadowsocks server port. Finally, we use Selenium~\cite{selenium} to navigate to each of these 1000 hostnames over HTTPS with and without Shadowsocks. Finally, we task each program: \name, Zeek~\cite{bro}, and Scapy~\cite{rohith2018scapy} to identify Shadowsocks proxy traffic using the algorithm described by Wu at al.~\cite{Wu2023a}. The results of this experiment are shown in Table~\ref{tab:performance_comparison}. In addition to the performance enhancements, the \name programs were more easier to implement than the Zeek equivalent, shown in Appendix~\ref{sec:zeek_shadowsocks}, which lacks key features required for implementing this classification scheme, and requires unwieldy implementations.

\subsection{ML-Based Censorship}\label{sec:ml_classification}
A major goal of \name, Goal \#3 as defined in Section~\ref{sec:ml_goal}, is to support ML-based censorship. ML-based censorship is the use of an ML model as part of the decision function that decides whether a connection is censored or not. Usually, these models are proposed to detect the \emph{behaviors} of circumvention protocols, as circumvention protocols often appear to passive observers with entirely benign packet-level metadata~\cite{domainfronting-fifield2015blocking} or use randomization to evade classification as a specific protocol~\cite{scramblesuit-10.1145/2517840.2517856}. A major example is the detection of Tor pluggable transports~\cite{wang2015seeing}. Many proposed ML detection techniques exploit weaknesses in the statistical patterns of circumvention protocols in order to promote efforts to make these protocols undetectable. There is no record of these techniques being used in a real-world setting, but an important aspect of \name is that it enables the emulation of hypothetical censors as well as real-world censors.

The fundamental building block of ML-based censorship is the classification model, as many problems in internet censorship such as protocol classification and website fingerprinting are essentially classification problems. \name is designed to easily emulate a hypothetical ML censor using such a model. There are two components to this: feature extraction, which computes features as packets are processed, and the model.




\subsubsection{ML Protocol Classification}


Multiple ML attacks against behavioral metadata of circumvention protocols have been proposed~\cite{wang2015seeing,xue2024FingerprintingObfuscatedProxy}, yet there has been no known application of these attacks by real-world censors. However, evaluating current circumvention strategies against these methods is useful in the case that such a strategy were to be deployed.

We  showcase \name's ability to emulate ML protocol classification by  emulating the models described by Wang et al~\cite{wang2015seeing}. First, convert the model trained by Wang et al. to ONNX format. Next, we create censor programs to extract and store the features required by the model, executing the model once the appropriate number of inputs has been reached. Finally, we evaluate these programs on captured PCAPs of web traffic and Obfs4 traffic, using the same data gathering method described in Section~\ref{sec:protocol_censorship} The results of this experiment are shown in Table~\ref{tab:performance_comparison}.

\subsubsection{Model Extraction}
\begin{figure}[t!]
    \centering
    \includegraphics[width=\columnwidth]{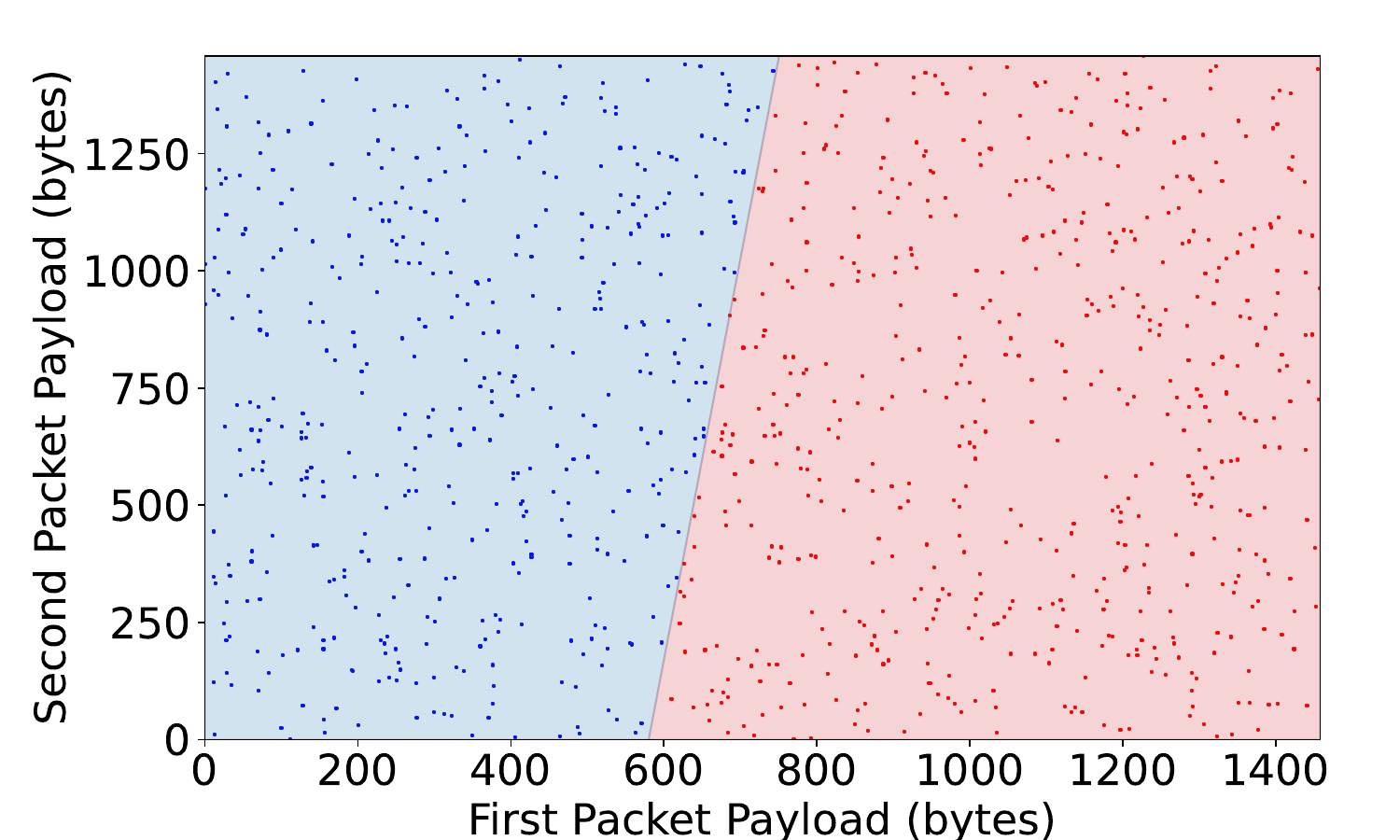}
    \caption{Each dot represents a censorship test using \name, and the background indicates the true decision boundary as stored by the model. Red is obfs4, while blue is web traffic.}
    \label{fig:model_extraction}
\end{figure}

As another ML case study, we show how  \name can be used to emulate censors for experimentation involving model extraction attacks~\cite{gong2020model} against  ML-enabled censors.
Specifically, we consider a scenario where the censors are using an ML model (e.g., a traffic classifier) for making their censorship decisions.
\name can be used to devise experiments towards inferring the ML model used by the censors (therefore, enabling the design of effective circumvention). 
%
In black-box model extraction, an attacker uses the publicly-exposed inference interface of a model to reconstruct a local version of that model. 
Because \name allows   loading  arbitrary models, and can provide feedback to clients on the network it censors, it is capable of properly emulating a black-box ML censor, against which model extraction attacks can be executed.

We demonstrate this by  performing a basic model extraction experiment. First, we train a linear regression model to predict HTTPS vs.\ obfs4 traffic, using the length of the first 2 non-empty packets. Next, we convert that model to ONNX, and run the censor program listed in Appendix~\ref{appendix:model_extraction_censor}. Then, we run \name in Tap mode on our client. Finally, we initiate 10000 connections from our client machine to a server, filling the first 2 data packets with random data of random lengths between 1 and 1480. The server is programmed to record which packets arrive using a similar methodology to protocol-based probing experiments~\cite{10.1145/3419394.3423644}. The results of this experiment are shown in Figure~\ref{fig:model_extraction}. In this figure, the background colors represent the true decision boundaries, while each dot represents a point of data learned by the model extraction program.


\subsubsection{Poisoning}

Model poisoning is the introduction of deliberately crafted data into a training dataset in order to trigger some outcome. For example, a censored user may wish to poison a censor's model to insert a backdoor allowing censorship to be evaded. For a censored user to have the opportunity to perform this attack, it is assumed that the censor is both using an ML model and frequently retraining their model using data captured from live traffic to known forbidden hosts, such as proxies.

The motivation for a censor to adopt this retraining regimen is that traffic data from circumvention protocols used by actual users is generally higher quality than synthetic traffic data~\cite{online-wf-277132}. The poisoning scenario is hypothetical, as no censor is currently known to be using ML or this sort of retraining method. However, development of poisoning techniques could be useful, as they would provide a powerful signal to indicate the presence of a frequently-retrained ML censor.

Implementing a poisoning scenario in \name first involves the logging of traffic to a specific proxy, which can be done outside of \name using TCPDUMP~\cite{tcpdump}
, or within \name using its logging facilities. With a selection of proxy traffic, the experimenter can train a model outside of \name using the capture data, then load it into \name using the method specified in Section~\ref{sec:ml_classification}. A client on a \name-hosted network then introduces deliberately crafted poison traffic. After re-training the model including the poison traffic, the client then re-sends this deliberately crafted traffic, and tests whether it matches the desired reaction. If so, the poisoning technique may be effective.

We perform an experiment to demonstrate this scenario using \name in order to showcase its ability to emulate advanced ML censorship scenarios. First, we train a model to classify HTTPS vs OBFS4 based only on the length and entropy of the first 30 packets. Then, we craft poison traffic~\cite{chen2017targeted} to introduce a backdoor allowing censorship evasion into the model, and add it into the training set. After retraining, we confirm the functionality of our poison sample.



\section{Evaluations}\label{sec:evaluations}
\begin{table*}[]
    \centering
    \begin{tabular}{c|c|c|c|c|c|c|}
    \toprule
                & \multicolumn{6}{c|}{Protocol}\\
    \midrule
         Censor & \hyperlink{data-http}{HTTP} & \hyperlink{data-dns}{DNS} & \hyperlink{data-https}{HTTPS} & \hyperlink{data-tor}{Tor} & \hyperlink{data-shadowsocks}{Shadowsocks} & \hyperlink{data-obfs4}{Obfs4} \\ 
    \midrule
    \hyperlink{censor-http-keyword}{HTTP keyword detection} & 0.996 & & & & &\\
    \hyperlink{censor-dns-injection}{DNS injection} & & 1.0 & & & &\\
    \hyperlink{censor-sni-filtering}{SNI filtering} & & & 0.93 & 0.5 & & \\
    \hyperlink{censor-gfw-encrypted}{GFW Shadowsocks detection}~\cite{Wu2023a} & & & & & 0.61 & 0.56 \\
    \bottomrule
    \end{tabular}
    \caption{Matrix demonstrating the effectiveness of various censorship strategies against various plaintext and obfuscated protocols.}
    \label{tab:censor-evaluation}
\end{table*}
In this section, we perform experiments to evaluate the use of \name for censorship circumvention research. We evaluate a number of network protocols and circumvention strategies, described in Section~\ref{sec:evaluation-data} against various censor strategies, described in Section~\ref{sec:evaluation-censors}. The result of these evaluations is shown in Table~\ref{tab:censor-evaluation}. Each experiment in Table~\ref{tab:censor-evaluation} corresponds to a live experiment run using \name. The resulting accuracies represent the percentage of successful \emph{allowed} connections averaged with the percentage of failed \emph{forbidden} connections. 

\subsection{Running Experiments in CensorLab}
A full experiment in \name is run as follows:
\name provides a harness for running censorship experiments (e.g. determining the efficacy of a particular strategy on a particular protocol) given two components provided by the user: a censor program and a client script. The censor program directs the censor, while the client program attempts to access network resources. We provide scripts to automatically provision a virtual machine, initialize \name, and run a client script. Client scripts may log successes and failures to a file, which is made available outside the virtual machine for analysis after the experiment has been run. By using a virtual machine for experiments, we are able to isolate experiments.

\subsection{Data}\label{sec:evaluation-data}
Each protocol consists of an ``allowed'' and ``forbidden'' subprotocol. These subprotocols are either variants of the same protocol, or a censorship circumvention protocol and an appropriate comparison protocol.

\hypertarget{data-http}{\noindent \textbf{HTTP}}: 
For both \emph{allowed} and \emph{forbidden} HTTP traffic, we perform 10000 \texttt{GET} requests to a remotely hosted HTTP server using the Python \texttt{requests}~\cite{2024PsfRequests} library. These requests are performed asynchronously, and thus with some concurrency. \emph{Forbidden} HTTP traffic contains GET requests with 5 randomly chosen words, hyphenated words. Of these words, one is is chosen from a list of words historically censored by the GFW~\cite{ng2024JasonqngChinesekeywords}. An example of this is \texttt{GET /follow-beach-authority-spare-ultrasurf}). The server is configured to respond with 1000 random words, with 20\% of words replaced with words chosen from the GFW censored set. \emph{Allowed} HTTP traffic is produced similarly, except all words are chosen such that they are not in the forbidden set.
\hypertarget{data-dns}{\noindent \textbf{DNS}}: 
To model a DNS censorship scenario, we perform 5000 \texttt{A} and 5000 \texttt{AAAA} DNS lookups. For the \emph{forbidden} dataset, we use domains censored by the GFW~\cite{hoang2021HowGreatGreata}. For the \emph{allowed} dataset, we use domains from the Tranco list~\cite{tranco-lepochat2019}, with domains from the forbidden set removed. These DNS lookups are performed asynchronously using the \texttt{trust-dns} library.

\hypertarget{data-https}{\noindent \textbf{HTTPS}}: 
To model HTTPS censorship of various types, we perform 1000 allowed and 1000 forbidden HTTPS GET requests using the domains in the \hyperlink{data-http}{HTTP Experiment}. These requests are performed using three methods, for a total of 3000 allowed requests and 300 forbidden requests. The methods are include browsing with Chrome, Firefox, and making requests using the Python requests~\cite{2024PsfRequests} library. For browser requests, Selenium is used for automation, and exceptions are used to detect errors. We use a variety of clients in order to test TLS fingerprinting strategies.

\noindent \hypertarget{data-tor}{\noindent \textbf{Tor}}: 
For the forbidden Tor dataset, we use the Tor proxy with Python's requests library to perform both HTTP and HTTPS get requests to the first 1000 domains in the Tranco~\cite{tranco-lepochat2019} list. Because Tor traffic contains some TLS headers, we use the allowed TLS dataset from the \hyperlink{data-http}{HTTP Experiment} as the allowed set.

\noindent \hypertarget{data-shadowsocks}{\noindent \textbf{Shadowsocks}}: 
For the forbidden Shadowsocks dataset, we use the same request strategy as in the \hyperlink{data-http}{HTTP Experiment}. We triple this process by using shadowsocks-go, shadowsocks-libev, and shadowsocks-rust as proxies. Because Shadowsocks aims to blend in with long-tail traffic, we use TII-SSRC-23~\cite{herzalla2023tii} as the corresponding allowed dataset. 

\noindent \hypertarget{data-obfs4}{\noindent \textbf{Obfs4}}: 
Obfs4 uses the same allowed dataset as Shadowsocks, as it is a fully-encrypted randomization protocol. The data collection for the forbidden dataset is identical to the Tor dataset except using Obfs4.


\subsection{Censors}\label{sec:evaluation-censors}

\hypertarget{censor-http-keyword}{\noindent \textbf{HTTP Keyword Detection}}\\
The HTTP keyword detection scenario simply checks the payload for the presence of keywords censored by the GFW.
The censor for this scenario is
\begin{minted}{python}
r = Regex("word1|word2|...")
def process(packet: Packet):
    if tcp.port == 80 and r.match(packet):
        return RESET
\end{minted}
\texttt{word1|word2|...} are words chosen from words censored by the GFW~\cite{ng2024JasonqngChinesekeywords}.

\noindent \hypertarget{censor-dns-injection}{\textbf{DNS Injection}}:
The DNS injection scenario uses the censor strategy specified in Section~\ref{sec:dns_showcase} where the list of domains is replaced with domains historically censored by the GFW~\cite{hoang2021HowGreatGreata}

\noindent \hypertarget{censor-sni-filtering}{\textbf{SNI filtering}}:
The SNI filtering scenario uses the censor program specified in Section~\ref{sec:keyword}, with the domains changed to those historically censored by the GFW~\cite{hoang2021HowGreatGreata}.

\noindent \hypertarget{censor-gfw-encrypted}{\textbf{GFW Shadowsocks Detection}}:
The GFW Shadowsocks detection censor uses the censor program described in Section~\ref{sec:protocol_censorship}.


\section{Ethical Considerations}\label{sec:ethics}

A potential ethical concern surrounding any censorship emulation platform like \name is whether they can be used by the censors to aid their development  of censorship mechanisms.   
However, censoring entities (e.g., nation states) already have significant resources and platforms that they use to  develop their censorship strategies. 
In particular, China's budget for censorship is estimated to be \emph{two orders of magnitude} higher than the overall budget available to the circumvention community, e.g., 6.6 billion dollars in 2020~\cite{fedasiuk2021buying}!  Very likely, these resourceful censors already have advanced \emph{proprietary} platforms and testbeds (akin to \name) that they use to test and develop censorship techniques. \name aims to help the circumvention community by reducing the time to react to new censorship techniques. The argument that \name might be used by the censors to harm Internet freedom is analogous to the infamous argument that privacy-enhancing technologies like  Tor may be used by (resourceful, skilled) cybercriminals.

\section{Conclusion}\label{sec:concl}
\name supports emulation of nearly every known deployed or hypothetical censorship  strategy, from basic strategies like keyword blocking to advanced strategies such as ML models requiring both aggregated and per-packet features.
\name is extensible, and can load arbitrary censorship routines, models, and configuration changes both at initialization and at runtime.
By combining Censor Programs, ML models, allow/blocklists, and censorship action patterns, \name provides a framework not just for experimentation, but for building a comprehensive collection of past and future censorship strategies. Such a collection would be invaluable to the developers of anti-censorship and traffic obfuscation tools, as they would be able to evaluate their techniques against a massive repertoire of countermeasures. We list censor programs that we have written emulating various censorship strategies in Table~\ref{tab:censorship_db}.

\section*{Acknowledgements}
This work was supported in part by NSF grants 2333965 and 1953786, and by the Young Faculty Award program of the Defense Advanced Research Projects Agency (DARPA) under the grant DARPA-RA-21-03-09-YFA9-FP-003.

\bibliographystyle{plain}
\bibliography{references}

\begin{thebibliography}{10}

\bibitem{tcpdump}
The Tcpdump~Group ~.
\newblock {{TCPDUMP}} \& {{LIBPCAP}}.
\newblock The Tcpdump Group, February 2024.

\bibitem{ieee-ethernet}
{802.3 WG - Ethernet Working Group}.
\newblock 802.3-2022 - {{IEEE Standard}} for {{Ethernet}}, July 2022.

\bibitem{tensorflow2015-whitepaper}
Mart\'{i}n Abadi, Ashish Agarwal, Paul Barham, Eugene Brevdo, Zhifeng Chen,
  Craig Citro, Greg~S. Corrado, Andy Davis, Jeffrey Dean, Matthieu Devin,
  Sanjay Ghemawat, Ian Goodfellow, Andrew Harp, Geoffrey Irving, Michael Isard,
  Yangqing Jia, Rafal Jozefowicz, Lukasz Kaiser, Manjunath Kudlur, Josh
  Levenberg, Dandelion Man\'{e}, Rajat Monga, Sherry Moore, Derek Murray, Chris
  Olah, Mike Schuster, Jonathon Shlens, Benoit Steiner, Ilya Sutskever, Kunal
  Talwar, Paul Tucker, Vincent Vanhoucke, Vijay Vasudevan, Fernanda Vi\'{e}gas,
  Oriol Vinyals, Pete Warden, Martin Wattenberg, Martin Wicke, Yuan Yu, and
  Xiaoqiang Zheng.
\newblock {TensorFlow}: Large-scale machine learning on heterogeneous systems,
  2015.
\newblock Software available from tensorflow.org.

\bibitem{10.1145/3419394.3423644}
Alice, Bob, Carol, Jan Beznazwy, and Amir Houmansadr.
\newblock How china detects and blocks shadowsocks.
\newblock In {\em Proceedings of the ACM Internet Measurement Conference}, IMC
  '20, page 111–124, New York, NY, USA, 2020. Association for Computing
  Machinery.

\bibitem{arduino}
{Arduino}.
\newblock Arduino, February 2024.

\bibitem{Attarian-Abdi-Hashemi-2019}
Reyhane Attarian, Lida Abdi, and Sattar Hashemi.
\newblock Adawfpa: Adaptive online website fingerprinting attack for tor
  anonymous network: A stream-wise paradigm.
\newblock {\em Computer Communications}, 148:74–85, Dec 2019.

\bibitem{bai2019}
Junjie Bai, Fang Lu, Ke~Zhang, et~al.
\newblock Onnx: Open neural network exchange.
\newblock \url{https://github.com/onnx/onnx}, 2019.

\bibitem{jafar}
Simone Basso.
\newblock Ooni/jafar.
\newblock Open Observatory of Network Interference (OONI), January 2023.

\bibitem{scapy-performance}
Philippe Biondi and Scapy Community.
\newblock Usage {\textemdash} {{Scapy}} 2.5.0 documentation, June 2023.

\bibitem{geneva-Bock2019a}
Kevin Bock, George Hughey, Xiao Qiang, and Dave Levin.
\newblock Geneva: {{Evolving Censorship Evasion Strategies}}.
\newblock In {\em Proceedings of the 2019 {{ACM SIGSAC Conference}} on
  {{Computer}} and {{Communications Security}}}, {{CCS}} '19, pages 2199--2214,
  {New York, NY, USA}, November 2019. {Association for Computing Machinery}.

\bibitem{bosshart2014P4}
Pat Bosshart, Dan Daly, Glen Gibb, Martin Izzard, Nick McKeown, Jennifer
  Rexford, Cole Schlesinger, Dan Talayco, Amin Vahdat, George Varghese, and
  David Walker.
\newblock P4: Programming protocol-independent packet processors.
\newblock {\em SIGCOMM Comput. Commun. Rev.}, 44(3):87--95, July 2014.

\bibitem{chai2019a}
Zimo Chai, Amirhossein Ghafari, and Amir Houmansadr.
\newblock On the {{Importance}} of {{Encrypted-SNI}} ({{ESNI}}) to {{Censorship
  Circumvention}}.
\newblock In {\em 9th {{USENIX}} Workshop on Free and Open Communications on
  the Internet ({{FOCI}} 19)}, pages 1--8, {Santa Clara, CA}, 2019. {USENIX
  Association}.

\bibitem{chen2017targeted}
Xinyun Chen, Chang Liu, Bo~Li, Kimberly Lu, and Dawn Song.
\newblock Targeted backdoor attacks on deep learning systems using data
  poisoning, 2017.

\bibitem{online-wf-277132}
Giovanni Cherubin, Rob Jansen, and Carmela Troncoso.
\newblock Online website fingerprinting: Evaluating website fingerprinting
  attacks on tor in the real world.
\newblock In {\em 31st USENIX Security Symposium (USENIX Security 22)}, pages
  753--770, Boston, MA, August 2022. USENIX Association.

\bibitem{selenium}
Software~Freedom Conservancy.
\newblock Selenium, 2023.

\bibitem{shadowsocks}
Shadowsocks Contributors.
\newblock Shadowsocks | a fast tunnel proxy that helps you bypass firewalls.
\newblock \url{https://shadowsocks.org/}, 2024.

\bibitem{tor-269582}
Roger Dingledine, Nick Mathewson, and Paul Syverson.
\newblock Tor: {{The Second-Generation Onion Router}}.
\newblock In {\em 13th {{USENIX Security Symposium}} ({{USENIX Security}} 04)},
  pages 1--18, {San Diego, CA}, August 2004. {USENIX Association}.

\bibitem{dixon2016network}
Lucas Dixon, Thomas Ristenpart, and Thomas Shrimpton.
\newblock Network traffic obfuscation and automated internet censorship.
\newblock {\em IEEE Security \& Privacy}, 14(6):43--53, 2016.

\bibitem{donenfeld2017wireguard}
Jason~A. Donenfeld.
\newblock {{WireGuard}}: {{Next Generation Kernel Network Tunnel}}.
\newblock In {\em Proceedings 2017 {{Network}} and {{Distributed System
  Security Symposium}}}, page~12, {San Diego, CA}, 2017. {Internet Society}.

\bibitem{el2017review}
Mohammed {El-Abd}.
\newblock A {{Review}} of {{Embedded Systems Education}} in the {{Arduino
  Age}}: {{Lessons Learned}} and {{Future Directions}}.
\newblock {\em International Journal of Engineering Pedagogy (iJEP)},
  7(2):79--93, May 2017.

\bibitem{fedasiuk2021buying}
Ryan Fedasiuk.
\newblock Buying silence: The price of internet censorship in china.
\newblock {\em China Brief}, 21(1):18--25, 2021.

\bibitem{domainfronting-fifield2015blocking}
David Fifield, Chang Lan, Rod Hynes, Percy Wegmann, and Vern Paxson.
\newblock Blocking-resistant communication through domain fronting.
\newblock {\em Proc. Priv. Enhancing Technol.}, 2015(2):46--64, 2015.

\bibitem{ooni}
Arturo Filast{\`o} and Jacob Appelbaum.
\newblock {{OONI}}: {{Open Observatory}} of {{Network Interference}}.
\newblock In {\em 2nd {{USENIX Workshop}} on {{Free}} and {{Open
  Communications}} on the {{Internet}} ({{FOCI}} 12)}, page~8, {Bellevue, WA},
  2012. {USENIX Association}.

\bibitem{gong2020model}
Xueluan Gong, Qian Wang, Yanjiao Chen, Wang Yang, and Xinchang Jiang.
\newblock Model extraction attacks and defenses on cloud-based machine learning
  models.
\newblock {\em IEEE Communications Magazine}, 58(12):83--89, 2020.

\bibitem{gulabovska2019survey}
Hristina Gulabovska and Zolt\{{\textbackslash}'a\}n\vphantom\{\}
  Porkol\{{\textbackslash}'a\}b.
\newblock Survey on {{Static Analysis Tools}} of {{Python Programs}}.
\newblock In {\em Proceedings of the {{Eighth Workshop}} on {{Software Quality
  Analysis}}, {{Monitoring}}, {{Improvement}}, and {{Applications}}}, page~10,
  {Ohrid, North Macedonia}, September 2019. {SQAMIA}.

\bibitem{hayes-danezis}
Jamie Hayes and George Danezis.
\newblock K-fingerprinting: A {{Robust Scalable Website Fingerprinting
  Technique}}.
\newblock In {\em 25th {{USENIX}} Security Symposium ({{USENIX}} Security 16)},
  pages 1187--1203, {Austin, TX}, August 2016. {USENIX Association}.

\bibitem{hayes2016k}
Jamie Hayes and George Danezis.
\newblock k-fingerprinting: A robust scalable website fingerprinting technique.
\newblock In {\em 25th USENIX Security Symposium (USENIX Security 16)}, pages
  1187--1203, Austin, TX, August 2016. USENIX Association.

\bibitem{herzalla2023tii}
Dania Herzalla, Willian~T Lunardi, and Martin Andreoni.
\newblock Tii-ssrc-23 dataset: Typological exploration of diverse traffic
  patterns for intrusion detection.
\newblock {\em IEEE Access}, 2023.

\bibitem{mitmproxy}
Maximilian Hils, Aldo Cortesi, and Thomas Kriechbaumer.
\newblock Mitmproxy/mitmproxy.
\newblock mitmproxy, February 2024.

\bibitem{hoang2021HowGreatGreata}
Nguyen~Phong Hoang, Arian~Akhavan Niaki, Jakub Dalek, Jeffrey Knockel, Pellaeon
  Lin, Bill Marczak, Masashi {Crete-Nishihata}, Phillipa Gill, and Michalis
  Polychronakis.
\newblock How {{Great}} is the {{Great Firewall}}? {{Measuring China}}'s dns
  {{Censorship}}.
\newblock In {\em 30th {{USENIX Security Symposium}} ({{USENIX Security}} 21)},
  pages 3381--3398, 2021.

\bibitem{houmansadrNonBlindWatermarkingNetwork2014}
Amir Houmansadr, Negar Kiyavash, and Nikita Borisov.
\newblock Non-{{Blind Watermarking}} of {{Network Flows}}.
\newblock {\em IEEE/ACM Transactions on Networking}, 22(4):1232--1244, August
  2014.

\bibitem{ieee-8021q}
{IEEE Standards Association}.
\newblock {{IEEE Standard}} for {{Local}} and {{Metropolitan Area
  Networks--Bridges}} and {{Bridged Networks}}, 2022.

\bibitem{india-censorship-Katira2023a}
Divyank Katira, Gurshabad Grover, Kushagra Singh, and Varun Bansal.
\newblock {{CensorWatch}}: {{On}} the {{Implementation}} of {{Online
  Censorship}} in {{India}}.
\newblock In {\em Free and {{Open Communications}} on the {{Internet}}},
  {{FOCI}} 2023, page~12, {Lausanne, Switzerland}, 2023. {Proceedings on
  Privacy Enhancing Technologies}.

\bibitem{internet-human-rights}
David Kravets.
\newblock {U.N. Report Declares Internet Access a Human Right}.
\newblock \url{http://www.wired.com/2011/06/internet-a-human-right/}, June
  2011.
\newblock Online Article.

\bibitem{tranco-lepochat2019}
Victor Le~Pochat, Tom Van~Goethem, Samaneh Tajalizadehkhoob, Maciej
  Korczy{\'n}ski, and Wouter Joosen.
\newblock Tranco: {{A}} research-oriented top sites ranking hardened against
  manipulation.
\newblock In {\em Proceedings of the 26th Annual Network and Distributed System
  Security Symposium}, {{NDSS}} 2019, pages 1--15, {San Diego, California.},
  February 2019. {NDSS}.

\bibitem{Lorimer2021a}
Anna~Harbluk Lorimer, Lindsey Tulloch, Cecylia Bocovich, and Ian Goldberg.
\newblock {{OUStralopithecus}}: {{Overt User Simulation}} for {{Censorship
  Circumvention}}.
\newblock In {\em Proceedings of the 20th {{Workshop}} on {{Workshop}} on
  {{Privacy}} in the {{Electronic Society}}}, {{WPES}} '21, pages 137--150,
  {New York, NY, USA}, November 2021. {Association for Computing Machinery}.

\bibitem{rust}
Nicholas~D. Matsakis and Felix~S. Klock.
\newblock The rust language.
\newblock {\em Ada Lett.}, 34(3):103–104, oct 2014.

\bibitem{mckeown2008OpenFlowEnablingInnovation}
Nick McKeown, Tom Anderson, Hari Balakrishnan, Guru Parulkar, Larry Peterson,
  Jennifer Rexford, Scott Shenker, and Jonathan Turner.
\newblock {{OpenFlow}}: Enabling innovation in campus networks.
\newblock {\em SIGCOMM Comput. Commun. Rev.}, 38(2):69--74, March 2008.

\bibitem{nasr2018deepcorr}
Milad Nasr, Alireza Bahramali, and Amir Houmansadr.
\newblock Deepcorr: Strong flow correlation attacks on tor using deep learning.
\newblock In {\em Proceedings of the 2018 ACM SIGSAC Conference on Computer and
  Communications Security}, CCS '18, page 1962–1976, New York, NY, USA, 2018.
  Association for Computing Machinery.

\bibitem{ng2024JasonqngChinesekeywords}
Jason~Q. Ng.
\newblock Jasonqng/chinese-keywords, March 2024.

\bibitem{iclab-9152784}
Arian~Akhavan Niaki, Shinyoung Cho, Zachary Weinberg, Nguyen~Phong Hoang, Abbas
  Razaghpanah, Nicolas Christin, and Phillipa Gill.
\newblock {{ICLab}}: {{A Global}}, {{Longitudinal Internet Censorship
  Measurement Platform}}.
\newblock In {\em 2020 {{IEEE Symposium}} on {{Security}} and {{Privacy}}
  ({{SP}})}, pages 135--151, {San Francisco, CA, USA}, May 2020. {IEEE}.

\bibitem{oh2022deepcoffea}
Se~Eun Oh, Taiji Yang, Nate Mathews, James~K Holland, Mohammad~Saidur Rahman,
  Nicholas Hopper, and Matthew Wright.
\newblock {{DeepCoFFEA}}: {{Improved Flow Correlation Attacks}} on {{Tor}} via
  {{Metric Learning}} and {{Amplification}}.
\newblock In {\em 2022 {{IEEE Symposium}} on {{Security}} and {{Privacy}}
  ({{SP}})}, pages 1915--1932, {San Francisco, CA, USA}, May 2022. {IEEE}.

\bibitem{ethical-measurement}
Craig Partridge and Mark Allman.
\newblock Ethical considerations in network measurement papers.
\newblock {\em Commun. ACM}, 59(10):58–64, sep 2016.

\bibitem{neurips2019-9015}
Adam Paszke, Sam Gross, Francisco Massa, Adam Lerer, James Bradbury, Gregory
  Chanan, Trevor Killeen, Zeming Lin, Natalia Gimelshein, Luca Antiga, Alban
  Desmaison, Andreas Kopf, Edward Yang, Zachary DeVito, Martin Raison, Alykhan
  Tejani, Sasank Chilamkurthy, Benoit Steiner, Lu~Fang, Junjie Bai, and Soumith
  Chintala.
\newblock {{PyTorch}}: {{An Imperative Style}}, {{High-Performance Deep
  Learning Library}}.
\newblock In {\em Advances in {{Neural Information Processing Systems}}},
  volume~32 of {\em {{NIPS}} 2019}, page~12, {Toronto, Canada}, 2019. {Curran
  Associates, Inc.}

\bibitem{bro}
Vern Paxson.
\newblock {Bro: a System for Detecting Network Intruders in Real-Time}.
\newblock {\em Computer Networks}, 31(23-24):2435--2463, 1999.

\bibitem{augur-7958591}
Paul Pearce, Roya Ensafi, Frank Li, Nick Feamster, and Vern Paxson.
\newblock Augur: {{Internet-Wide Detection}} of {{Connectivity Disruptions}}.
\newblock In {\em 2017 {{IEEE Symposium}} on {{Security}} and {{Privacy}}
  ({{SP}})}, pages 427--443, {San Jose, California}, May 2017. {IEEE}.

\bibitem{scikit-learn}
Fabian Pedregosa, Gael Varoquaux, Alexandre Gramfort, Vincent Michel, Bertrand
  Thirion, Olivier Grisel, Mathieu Blondel, Peter Prettenhofer, Ron Weiss,
  Vincent Dubourg, Jake Vanderplas, Alexandre Passos, David Cournapeau,
  Matthieu Brucher, Matthieu Perrot, and Edouard Duchesnay.
\newblock Scikit-learn: Machine learning in {P}ython.
\newblock {\em Journal of Machine Learning Research}, 12:2825--2830, 2011.

\bibitem{ooni-probe}
OONI Project.
\newblock Ooni probe.
\newblock \url{https://ooni.org/install/}, Jan 2022.

\bibitem{iptables}
The~Netfilter Project.
\newblock netfilter/iptables project homepage - the netfilter.org
  “nftables” project, Oct 2011.

\bibitem{snowflake}
The~Tor Project.
\newblock Snowflake.
\newblock \url{https://snowflake.torproject.org/}, 2018.

\bibitem{2024PsfRequests}
{Python Software Foundation}.
\newblock Psf/requests.
\newblock Python Software Foundation, July 2024.

\bibitem{middleboxes-Raman2022a}
Ram~Sundara Raman, Mona Wang, Jakub Dalek, Jonathan Mayer, and Roya Ensafi.
\newblock Network measurement methods for locating and examining censorship
  devices.
\newblock In {\em Proceedings of the 18th {{International Conference}} on
  Emerging {{Networking EXperiments}} and {{Technologies}}}, {{CoNEXT}} '22,
  pages 18--34, {New York, NY, USA}, November 2022. {Association for Computing
  Machinery}.

\bibitem{v2ray}
Darien Raymond and Xiaokang Wang.
\newblock V2ray/v2ray-core.
\newblock Project V, February 2024.

\bibitem{Rimmer-Preuveneers-Juarez-Goethem-Joosen-2018}
Vera Rimmer, Davy Preuveneers, Marc Juarez, Tom~Van Goethem, and Wouter Joosen.
\newblock Automated {{Website Fingerprinting}} through {{Deep Learning}}.
\newblock In {\em Proceedings 2018 {{Network}} and {{Distributed System
  Security Symposium}}}, page~15, {San Diego, CA}, 2018. {Internet Society}.

\bibitem{rustp}
{RustPython Project}.
\newblock Rustpython.
\newblock \url{https://github.com/RustPython/RustPython}, 2023.

\bibitem{rohith2018scapy}
Rohith~Raj S, Rohith R, Minal Moharir, and Shobha G.
\newblock Scapy- a powerful interactive packet manipulation program.
\newblock In {\em 2018 {{International Conference}} on {{Networking}},
  {{Embedded}} and {{Wireless Systems}} ({{ICNEWS}})}, pages 1--5, {Bangalore,
  India}, December 2018. {IEEE}.

\bibitem{sakamoto2024BleedingWallHematologic}
{Sakamoto} and Elson Wedwards.
\newblock Bleeding {{Wall}}: {{A Hematologic Examination}} on the {{Great
  Firewall}}.
\newblock In {\em Free and {{Open Communications}} on the {{Internet}}},
  page~9, {Online}, 2024. {Proceedings on Privacy Enhancing Technologies}.

\bibitem{salesforce2024SalesforceJa3}
{Salesforce}.
\newblock Salesforce/ja3.
\newblock Salesforce, March 2024.

\bibitem{schaller1997moore}
Robert~R Schaller.
\newblock Moore's law: past, present and future.
\newblock {\em IEEE spectrum}, 34(6):52--59, 1997.

\bibitem{Sharma2021a}
Piyush~Kumar Sharma, Devashish Gosain, and Sambuddho Chakravarty.
\newblock Camoufler: {{Accessing The Censored Web By Utilizing Instant
  Messaging Channels}}.
\newblock In {\em Proceedings of the 2021 {{ACM Asia Conference}} on
  {{Computer}} and {{Communications Security}}}, {{ASIA CCS}} '21, pages
  147--161, {New York, NY, USA}, June 2021. {Association for Computing
  Machinery}.

\bibitem{Sirinam-Imani-Juarez-Wright-2018}
Payap Sirinam, Mohsen Imani, Marc Juarez, and Matthew Wright.
\newblock Deep {{Fingerprinting}}: {{Undermining Website Fingerprinting
  Defenses}} with {{Deep Learning}}.
\newblock In {\em Proceedings of the 2018 {{ACM SIGSAC Conference}} on
  {{Computer}} and {{Communications Security}}}, {{CCS}} '18, pages 1928--1943,
  {New York, NY, USA}, October 2018. {Association for Computing Machinery}.

\bibitem{triplet-fingerprinting}
Payap Sirinam, Nate Mathews, Mohammad~Saidur Rahman, and Matthew Wright.
\newblock Triplet fingerprinting: More practical and portable website
  fingerprinting with n-shot learning.
\newblock In {\em Proceedings of the 2019 ACM SIGSAC Conference on Computer and
  Communications Security}, CCS '19, page 1131–1148, New York, NY, USA, 2019.
  Association for Computing Machinery.

\bibitem{netcontrol}
Jon Siwek.
\newblock {{NetControl Framework}} {\textemdash} {{Book}} of {{Zeek}}
  (git/master), February 2024.

\bibitem{censoredplanet-10.1145/3372297.3417883}
Ram Sundara~Raman, Prerana Shenoy, Katharina Kohls, and Roya Ensafi.
\newblock Censored planet: An internet-wide, longitudinal censorship
  observatory.
\newblock In {\em Proceedings of the 2020 ACM SIGSAC Conference on Computer and
  Communications Security}, CCS '20, page 49–66, New York, NY, USA, 2020.
  Association for Computing Machinery.

\bibitem{tencent-cost}
Tencent.
\newblock Cloud virtual machine - pricing.
\newblock \url{https://www.tencentcloud.com/pricing/cvm/calculator}, 2023.

\bibitem{Ben2018Quack}
Benjamin VanderSloot, Allison McDonald, Will Scott, J.~Alex Halderman, and Roya
  Ensafi.
\newblock Quack: {{Scalable}} remote measurement of {{Application-Layer}}
  censorship.
\newblock In {\em 27th {{USENIX}} Security Symposium ({{USENIX}} Security 18)},
  pages 187--202, {Baltimore, MD}, August 2018. {USENIX Association}.

\bibitem{Verkamp2012a}
John-Paul Verkamp and Minaxi Gupta.
\newblock Inferring mechanics of web censorship around the world.
\newblock In {\em 2nd {{USENIX}} Workshop on Free and Open Communications on
  the Internet ({{FOCI}} 12)}, page~7, {Bellevue, WA}, August 2012. {USENIX
  Association}.

\bibitem{wang2015seeing}
Liang Wang, Kevin~P. Dyer, Aditya Akella, Thomas Ristenpart, and Thomas
  Shrimpton.
\newblock Seeing through {{Network-Protocol Obfuscation}}.
\newblock In {\em Proceedings of the 22nd {{ACM SIGSAC Conference}} on
  {{Computer}} and {{Communications Security}}}, pages 57--69, {Denver Colorado
  USA}, October 2015. {ACM}.

\bibitem{wangEffectiveAttacksProvable}
Tao Wang, Xiang Cai, Rishab Nithyanand, Rob Johnson, and Ian Goldberg.
\newblock Effective {{Attacks}} and {{Provable Defenses}} for {{Website
  Fingerprinting}}.
\newblock In {\em 23rd {{USENIX}} Security Symposium ({{USENIX}} Security 14)},
  pages 143--157, {San Diego, CA}, August 2014. {USENIX Association}.

\bibitem{wang2013improved}
Tao Wang and Ian Goldberg.
\newblock Improved website fingerprinting on {{Tor}}.
\newblock In {\em Proceedings of the 12th {{ACM}} Workshop on {{Workshop}} on
  Privacy in the Electronic Society}, pages 201--212, {Berlin Germany},
  November 2013. {ACM}.

\bibitem{Wang-Goldberg-2016}
Tao Wang and Ian Goldberg.
\newblock On realistically attacking tor with website fingerprinting.
\newblock {\em Proceedings on Privacy Enhancing Technologies}, 2016(4):21–36,
  Oct 2016.

\bibitem{winter2012great}
Philipp Winter and Stefan Lindskog.
\newblock How the {{Great Firewall}} of {{China}} is blocking {{Tor}}.
\newblock In {\em Free and Open Communications on the Internet}, page~7,
  {Bellevue, WA}, 2012. {USENIX}.

\bibitem{scramblesuit-10.1145/2517840.2517856}
Philipp Winter, Tobias Pulls, and Juergen Fuss.
\newblock Scramblesuit: A polymorphic network protocol to circumvent
  censorship.
\newblock In {\em Proceedings of the 12th ACM Workshop on Workshop on Privacy
  in the Electronic Society}, WPES '13, page 213–224, New York, NY, USA,
  2013. Association for Computing Machinery.

\bibitem{Wu2023a}
Mingshi Wu, Jackson Sippe, Danesh Sivakumar, Jack Burg, Peter Anderson,
  Xiaokang Wang, Kevin Bock, Amir Houmansadr, Dave Levin, and Eric Wustrow.
\newblock How the {{Great Firewall}} of {{China Detects}} and {{Blocks Fully
  Encrypted Traffic}}.
\newblock In {\em 32nd {{USENIX}} Security Symposium ({{USENIX}} Security 23)},
  pages 2653--2670, {Anaheim, CA}, August 2023. {USENIX Association}.

\bibitem{xue2024FingerprintingObfuscatedProxy}
Diwen Xue, Michalis Kallitsis, Amir Houmansadr, and Roya Ensafi.
\newblock Fingerprinting {{Obfuscated Proxy Traffic}} with {{Encapsulated TLS
  Handshakes}}.
\newblock In {\em Usenix {{Security Symposium}} 2024}, page~18, {Philadelphia,
  PA}, 2024. {USENIX Association}.

\bibitem{russia-censorship-xue2022b}
Diwen Xue, Benjamin {Mixon-Baca}, ValdikSS, Anna Ablove, Beau Kujath,
  Jedidiah~R. Crandall, and Roya Ensafi.
\newblock {{TSPU}}: {{Russia}}'s decentralized censorship system.
\newblock In {\em Proceedings of the 22nd {{ACM Internet Measurement
  Conference}}}, {{IMC}} '22, pages 179--194, {New York, NY, USA}, October
  2022. {Association for Computing Machinery}.

\bibitem{xue2022OpenVPNOpenVPN}
Diwen Xue, Reethika Ramesh, Arham Jain, Michalis Kallitsis, J.~Alex Halderman,
  Jedidiah~R. Crandall, and Roya Ensafi.
\newblock \{\vphantom\}{{OpenVPN}}\vphantom\{\} is {{Open}} to
  \{\vphantom\}{{VPN}}\vphantom\{\} {{Fingerprinting}}.
\newblock In {\em 31st {{USENIX Security Symposium}} ({{USENIX Security}} 22)},
  pages 483--500, {Boston, MA}, 2022. {USENIX Association}.

\bibitem{xue2021throttling}
Diwen Xue, Reethika Ramesh, Valdik~S S, Leonid Evdokimov, Andrey Viktorov,
  Arham Jain, Eric Wustrow, Simone Basso, and Roya Ensafi.
\newblock Throttling {{Twitter}}: An emerging censorship technique in
  {{Russia}}.
\newblock In {\em Proceedings of the 21st {{ACM Internet Measurement
  Conference}}}, {{IMC}} '21, pages 435--443, {New York, NY, USA}, November
  2021. {Association for Computing Machinery}.

\end{thebibliography}

\appendix

\section*{Availability}
All code for \name will be made free and open source, with the intent to maintain and update its capabilities continuously as new censorship and censorship evasion methods emerge.

\section*{Overview of Circumvention Approaches}\label{sec:circumvention_approaches}

The most common class of circumvention system is are \emph{proxies}, which route traffic through another computer in order to conceal its true destination. However, basic proxies often fail to successfully evade censorship unless they introduce obfuscation mechanisms. A prominent technique used by circumvention tools is \textit{Tunneling}, which hides traffic inside traffic to either reputable domains~\cite{domainfronting-fifield2015blocking,snowflake} or volunteers~\cite{snowflake} in order to exploit a censor's unwillingness to inflict collateral damage against prominent services or potential customers, respectively. Some examples of tunneling protocols include Meek, which tunnels traffic over HTTPS using domain fronting~\cite{domainfronting-fifield2015blocking} and Snowflake~\cite{snowflake}, which uses WebRTC DTLS tunnels. Another popular class of circumvention techniques is \textit{Randomization}. Randomization techniques introduce noisy or random characteristics to the circumvention protocol, causing Deep Packet Inspection (DPI) middleboxes to treat the circumvention traffic as it would any unknown protocol~\cite{scramblesuit-10.1145/2517840.2517856}. Some popular randomization protocols are Obfs4~\cite{scramblesuit-10.1145/2517840.2517856}, ShadowSocks~\cite{shadowsocks}. Finally, a new class of circumvention protocol has recently emerged, which we refer to as \textit{fuzzing-based circumvention}. In fuzzing-based circumvention, rather than targeting broad flaws in DPI systems such as in \textit{randomization} protocols, the system automatically finds and exploits specific flaws in a deployed censorship system, such as protocol parsing or TCP aggregation. One example of such a system is Geneva~\cite{geneva-Bock2019a}, which utilizes a genetic fuzzing approach. 

Each class of circumvention suffers from a risk: if certain assumptions fail to hold, such as with a change in censor strategy, the circumvention will no longer work. Tunneling protocols rely on the avoidance of collateral damage, but may lose efficacy as a censor decides that an increased amount of collateral damage is acceptable. Randomization protocols have seen direct attacks, such as China's encrypted protocol detection system~\cite{Wu2023a}. Finally, fuzzing-based protocols show promise, but lack rigorous testing and may overfit to a given censor.

\section*{IPC Commands}\label{appendix:ipc_commands}
In this section, we list each IPC command supported by \name, using both the CLI and Python interfaces. For commands controlling the allowlist, each instance of \texttt{allowlist} may also be \texttt{blocklist}. We use \texttt{udp-service} as an example identifier, but each call also supports any identifier specified in Table~\ref{tab:identifiers}.
\begin{itemize}
    \item \textbf{Shutdown}: Shuts down \name\\
    \texttt{clipc shutdown}
    \begin{minted}{python}
censorlab.shutdown()
    \end{minted}
    \item \textbf{Reload}: Reloads \texttt{censor.toml} and clears all state\\
    \texttt{clipc reload}
    \begin{minted}{python}
censorlab.reload()
    \end{minted}
    \item \textbf{Allowlist action}: Changes the reject action for a allowlist\\
    \texttt{clipc allowlist action udp-service drop}
    \begin{minted}{python}
censorlab.allowlist.udp_service.action("drop")
    \end{minted}
    \item \textbf{Allowlist add}: Adds entries to the allowlist\\
    \texttt{clipc allowlist add udp-service 8.8.8.8 53}
    \begin{minted}{python}
censorlab.allowlist.udp_service.add(
    "8.8.8.8", 53
)
    \end{minted}
    \item \textbf{Allowlist remove}: Removes entries from the allowlist\\
    \texttt{clipc allowlist remove udp-service 8.8.8.8 53}
    \begin{minted}{python}
censorlab.allowlist.udp_service.remove(
    "8.8.8.8", 53
)
    \end{minted}
    \item \textbf{Allowlist list}: Lists entries in the allowlist\\
    \texttt{clipc allowlist list udp-service}
    \begin{minted}{python}
censorlab.allowlist.udp_service.list()
    \end{minted}
    \item \textbf{Program load}: Loads a new program. Clears all connection state\\
    \texttt{clipc program load python censor.py}
    \begin{minted}{python}
censorlab.program.load(
    "python", "censor.py"
)
    \end{minted}
    \item \textbf{Debug dump}: Dumps the program state for a given connection\\
    \texttt{clipc debug dump 192.168.1.2 8.8.8.8 23212 53 udp}
    \begin{minted}{python}
censorlab.debug.dump(
    "192.168.1.2","8.8.8.8","23212","53","udp"
)
    \end{minted}
    \item \textbf{Model add}:  loads an ONNX model and makes it available under the given name\\
    \texttt{clipc model add wf ./models/wf.onnx}
    \begin{minted}{python}
censorlab.model.add("wf", "./models/wf.onnx")
    \end{minted}
    \item \textbf{Model remove}:  unloads an ONNX model by the given name\\
    \texttt{clipc model remove wf}
    \begin{minted}{python}
censorlab.model.remove("wf")
    \end{minted}
\end{itemize}

\section*{CensorLang Grammar}\label{appendix:censorlang_grammar}

Figure~\ref{fig:censor_program_grammar} shows the grammar of CensorLang.
\begin{figure}[ht!]
    \centering
    \begin{bnf*}
      \bnfprod{program}{
               \bnfpn{line}
        \bnfor \bnfpn{line} \bnfsp \bnfts{;} \bnfsp \bnfpn{program}
      }\\
      \bnfprod{line} {
               \bnfts{if} \bnfpn{condition} \bnfts{:} \bnfpn{operation}
        \bnfor \bnfpn{operation}
      }\\
      \bnfprod{condition} {
            \bnfts{REGEX} \bnfor
            \bnfpn{input} \bnfpn{operator} \bnfpn{input}
      }\\
      \bnfprod{input} {
                   \bnfts{field} \bnfts{:} \bnfpn{field}
            \bnfor \bnfpn{register}
            \bnfor \bnfpn{float}
      }\\
      \bnfmore{
            \bnfor \bnfpn{int}
            \bnfor \bnfpn{bool}
      }\\
      \bnfprod{field} {
        \bnftd{name of a field, such as "payload.len"}
      }\\
      \bnfprod{register} {
        \bnfpn{register-class} \bnfts{:} \bnfpn{uint} \bnfts{.} \bnfpn{register-type}
      }\\
      \bnfprod{register-class} {
        \bnfts{reg} \bnfor \bnfts{src} \bnfor \bnfts{dst}
      }\\
      \bnfprod{register-type} {
        \bnfts{f32} \bnfor \bnfts{f64} \bnfor \bnfts{i32} \bnfor \bnfts{u32} \bnfor
      }\\
      \bnfmore{
         \bnfts{i64} \bnfor \bnfts{u64} \bnfor \bnfts{b}
      }\\
      \bnfprod{float} {
        \bnftd{A signed floating point number}
      }\\
      \bnfprod{int} {
        \bnftd{A signed integer}
      }\\
      \bnfprod{uint} {
        \bnftd{An unsigned integer}
      }\\
      \bnfprod{bool} {
        \bnfts{true} \bnfor \bnfts{false}
      }\\
      \bnfprod{operator} {
        \bnfts{<=} \bnfor
        \bnfts{<} \bnfor
        \bnfts{!=} \bnfor
        \bnfts{==} \bnfor
        \bnfts{>=} \bnfor
        \bnfts{>} \bnfor
      }\\
      \bnfmore{
        \bnfts{LEQ} \bnfor
        \bnfts{LT} \bnfor
        \bnfts{NEQ} \bnfor
        \bnfts{EQ} \bnfor
        \bnfts{GEQ} \bnfor
        \bnfts{GT} \bnfor
      }\\
      \bnfmore{
        \bnfts{and} \bnfor
        \bnfts{or} \bnfor
        \bnfts{xor}
      }\\
      \bnfprod{operation} {
        \bnfts{COPY} \bnfpn{input} \bnfts{->} \bnfpn{register} \bnfor
      }\\
      \bnfmore{
        \bnfts{ADD} \bnfpn{input} \bnfts{,} \bnfpn{input} \bnfts{->} \bnfpn{register} \bnfor
      }\\
      \bnfmore{
        \bnfts{SUB} \bnfpn{input} \bnfts{,} \bnfpn{input} \bnfts{->} \bnfpn{register} \bnfor
      }\\
      \bnfmore{
        \bnfts{MUL} \bnfpn{input} \bnfts{,} \bnfpn{input} \bnfts{->} \bnfpn{register} \bnfor
      }\\
      \bnfmore{
        \bnfts{DIV} \bnfpn{input} \bnfts{,} \bnfpn{input} \bnfts{->} \bnfpn{register} \bnfor
      }\\
      \bnfmore{
        \bnfts{MOD} \bnfpn{input} \bnfts{,} \bnfpn{input} \bnfts{->} \bnfpn{register} \bnfor
      }\\
      \bnfmore{
        \bnfts{AND} \bnfpn{input} \bnfts{,} \bnfpn{input} \bnfts{->} \bnfpn{register} \bnfor
      }\\
      \bnfmore{
        \bnfts{OR} \bnfpn{input} \bnfts{,} \bnfpn{input} \bnfts{->} \bnfpn{register} \bnfor
      }\\
      \bnfmore{
        \bnfts{XOR} \bnfpn{input} \bnfts{,} \bnfpn{input} \bnfts{->} \bnfpn{register} \bnfor
      }\\
      \bnfmore{
        \bnfts{RETURN} \bnfpn{action} \bnfor
      }\\
      \bnfmore{
        \bnfts{MODEL}
      }\\
      \bnfprod{action}{
        \bnfts{allow} \bnfor \bnfts{ignore} \bnfor \bnfts{drop} \bnfor \bnfts{reset}
      }\\
      \bnfmore{
        \bnfor \bnfts{delay} \bnfts{:} \bnfts{float}
      }
    \end{bnf*}
    \caption{Grammar of Censor Programs}
    \label{fig:censor_program_grammar}
\end{figure}

\section*{Showcase: Shadowsocks}
As shown in Section~\ref{sec:showcases}, \name is a much easier interface for programming the Shadowsocks detection task~\cite{Wu2023a}. In this section, we list code samples for both Zeek and PyCL in order to demonstrate this discrepancy.
\subsection{Zeek}\label{sec:zeek_shadowsocks}
\begin{minted}{bro}
event zeek_init() {

}
function is_printable( b: string ): bool {
  local i = bytestring_to_count(b);
  return i >= 0x20 && i <=0x7E; 
}
function popcount( b: string ): count  {
  local i = bytestring_to_count(b);
  local c = 0;
  if ( i >= 128 ) {
    c += 1;
  };
  if ( i >= 64 ) {
    c += 1;
  };
  if ( i >= 32 ) {
    c += 1;
  };
  if ( i >= 16 ) {
    c += 1;
  };
  if ( i >= 8 ) {
    c += 1;
  };
  if ( i >= 4 ) {
    c += 1;
  };
  if ( i >= 2 ) {
    c += 1;
  };
  if ( i >= 1 ) {
    c += 1;
  };
  return c;
}
event tcp_packet(
    c: connection,
    is_orig: bool,
    flags: string,
    seq: count,
    ack: count,
    len: count,
    payload: string
) {
  local pload = sub_bytes(payload, 0, len);
  local first_6 = T;
  for ( b in pload[:6] ) {
    if ( !is_printable(b) ){
      first_6 = F;
      break;
    } 
  }
  if ( first_6 ) {
    return;
  }
  local num_printable = 0;
  local popcount_total = 0; 
  local streak = 0;
  for ( b in pload ) {
    if ( is_printable(b) ) {
      num_printable += 1;
      streak += 1;
      if (streak > 20) {
        return;
      } 
    }
    else {
      streak = 0;
    }
    popcount_total += popcount(b);
  }
  if ( num_printable > (len/2) ) {
    return;
  }
  local popcount_normalized = (
    count_to_double(popcount_total) /
    count_to_double(len)
  );
  if ( popcount_total <= 3.4 ||
       popcount_total >= 4.6 ) {
    return;
  }
}
\end{minted}

\subsection{Scapy}
\begin{minted}{python}
import re
import sys
import time

from scapy.all import rdpcap
from scapy.layers.inet import TCP

start = time.time()
reg = re.compile(
  "^[\x20-\x7E]{6} | .*[\x20-\x7E]{20}.* |" +
  "^HTTP | ^[\x16-\x17]\x03[\x00-\x09]"
)
pcap_path = sys.argv[1]


def is_printable(b: int):
    return b >= 0x20 and b <= 0x7E


def process(packet):
    try:
        if TCP not in packet:
            return
        payload = packet[TCP].payload
        # Check first few bytes
        if reg.match(payload) is not None:
            return
        total_printable = 0
        # Check popcount
        p = 0
        for b in payload:
            p += b.bit_count()
        popcount_avg = p / len(payload)
        if popcount_avg <= 3.4 or popcount_avg >= 4.6:
            return
        return True
    except:
        pass


pcap = rdpcap(pcap_path)
for packet in pcap:
    if process(packet) is not None:
        print("Shadowsocks")
end = time.time()
elapsed = end - start
print(elapsed * 1000000)
\end{minted}
\section{Model Extraction Censor}\label{appendix:model_extraction_censor}
\begin{minted}{python}
idx = 0
lens = [0,0]
if packet.payload_len > 0 and idx < 2:
    lens[idx] = packet.payload_len
    idx += 1
    if idx == 2:
        if model(lens) > 0):
            return DROP
\end{minted}

\section*{Database of Censor Strategies}
Table~\ref{tab:censorship_db} lists our current database of censorship strategies for \name. A green check (\cmark) indicates scenarios for which censor programs are provided. An orange check (\maybecheck) indicates scenarios that may be supported by censor programs. A red X (\xmark) indicates scenarios that would require internal modifications to \name to support.

\begin{table*}[h]
\centering
\caption{List of censorship strategies currently implemented in \name's database.}\label{tab:censorship_db}
\begin{tabularx}{0.96\textwidth}{c|X|c}
\toprule
\multicolumn{3}{c}{Past and Current Strategies}\\
\midrule
Category & Description & Supported\\
\midrule
TCP Vulnerabilities & Censors with various buggy TCP aggregation schemes\cite{geneva-Bock2019a} & \cmark\\
DNS Vulnerabilities  & Buffer overflows in packet parsing~\cite{sakamoto2024BleedingWallHematologic} & \xmark \\
DNS Response Injection & Injects {\tt NXDOMAIN} on forbidden DNS requests & \cmark \\
DNS Response Injection & Injects null-routed IP or blockpage on forbidden DNS requests& \cmark\\
DNS Drop forbidden & Drops forbidden requests and responses& \cmark \\
HTTP RST forbidden & TCP RST based on forbidden keywords& \cmark \\
HTTP Drop forbidden & Drop based on on forbidden keywords& \cmark \\
HTTPS Censorship & SNI-based blocking & \cmark\\
HTTPS Censorship & Handshake fingerprinting (e.g. JA3~\cite{salesforce2024SalesforceJa3}) & \maybecheck\\
Shadowsocks & GFW probe trigger based on length and entropy discovered by Alice et al~\cite{10.1145/3419394.3423644}  & \cmark \\
Shadowsocks & GFW blocking based on TCP payload discovered by Wu et al~\cite{Wu2023a} & \cmark \\
VPN detection & Detection of OpenVPN~\cite{xue2022OpenVPNOpenVPN}, WireGuard~\cite{donenfeld2017wireguard}, etc & \maybecheck\\
Per-host state & Decisions based on per-host state rather than per-connection state& \xmark\\
\midrule
\multicolumn{3}{c}{Futuristic Strategies}\\
\midrule
ML - Website Fingerprinting & Using various algorithms~\cite{triplet-fingerprinting,wang2013improved,wangEffectiveAttacksProvable,hayes2016k} & \maybecheck\\
ML - Protocol classification & Blocking Obfs4/Meek~\cite{wang2015seeing} & \cmark\\
ML - Protocol classification & Detecting the use of encapsulated TLS handshakes~\cite{xue2024FingerprintingObfuscatedProxy} & \maybecheck\\
ML - Watermarking & Use active delays and modifications for flow watermarking~\cite{houmansadrNonBlindWatermarkingNetwork2014} & \maybecheck\\
\bottomrule
\end{tabularx}
\end{table*}

\end{document}